\documentclass[conference]{IEEEtran}
\IEEEoverridecommandlockouts
\usepackage{cite}
\usepackage{float}
\usepackage{url}
\usepackage{amsmath,amssymb,amsfonts}
\usepackage{algorithmic}
\usepackage{graphicx}
\usepackage{textcomp}
\usepackage{xcolor}
\usepackage{multirow}
\usepackage{booktabs}
\usepackage{tikz}
\usepackage{float}
\usepackage{subfig}
\usepackage{xspace}
\usepackage[numbers]{natbib}
\def\BibTeX{{\rm B\kern-.05em{\sc i\kern-.025em b}\kern-.08em
    T\kern-.1667em\lower.7ex\hbox{E}\kern-.125emX}}
\newcommand*\circled[1]{\tikz[baseline=(char.base)]{
            \node[shape=circle,draw,inner sep=2pt] (char) {#1};}}
\newcommand{\mfnn}{\emph{MF-CGNN}\xspace}

\begin{document}

\title{Comparative Study of Large Language Model Architectures on Frontier \\
\thanks{This manuscript has been co-authored by UT-Battelle, LLC, under contract DE-AC05-00OR22725 with the US Department of Energy (DOE). The US government retains and the publisher, by accepting the article for publication, acknowledges that the US government retains a nonexclusive, paid-up, irrevocable, worldwide license to publish or reproduce the published form of this manuscript, or allow others to do so, for US government purposes. DOE will provide public access to these results of federally sponsored research in accordance with the DOE Public Access Plan (http://energy.gov/downloads/doe-public-access-plan).}

}

\makeatletter
\newcommand{\linebreakand}{%
  \end{@IEEEauthorhalign}
  \hfill\mbox{}\par
  \mbox{}\hfill\begin{@IEEEauthorhalign}
}
\makeatother

\author{\IEEEauthorblockN{Junqi Yin}
\IEEEauthorblockA{\textit{Oak Ridge National Laboratory} \\
%\textit{Oak Ridge National Laboratory}\\
Oak Ridge, TN \\
yinj@ornl.gov}
\and
\IEEEauthorblockN{Avishek Bose}
\IEEEauthorblockA{\textit{Oak Ridge National Laboratory} \\
Oak Ridge, TN \\
bosea@ornl.gov}
\and
\IEEEauthorblockN{Guojing Cong}
\IEEEauthorblockA{\textit{Oak Ridge National Laboratory} \\
Oak Ridge, TN \\
congg@ornl.gov}
\linebreakand
\IEEEauthorblockN{Isaac Lyngaas}
\IEEEauthorblockA{\textit{Oak Ridge National Laboratory} \\
Oak Ridge, TN \\
lyngaasir@ornl.gov}
\and
\IEEEauthorblockN{Quentin Anthony}
\IEEEauthorblockA{\textit{Ohio State University} \\
\textit{EleutherAI}\\
Columbus, OH \\
qubitquentin@gmail.com}
}

\maketitle
%\IEEEpeerreviewmaketitle
\thispagestyle{empty}

\begin{abstract}
Large language models (LLMs) have garnered significant attention in both the AI community and beyond. Among these, the Generative Pre-trained Transformer (GPT) has emerged as the dominant architecture, spawning numerous variants. However, these variants have undergone pre-training under diverse conditions, including variations in input data, data preprocessing, and training methodologies, resulting in a lack of controlled comparative studies. Here we meticulously examine two prominent open-sourced GPT architectures, GPT-NeoX and LLaMA, leveraging the computational power of Frontier, the world's first Exascale supercomputer. Employing the same materials science text corpus and a comprehensive end-to-end pipeline, we conduct a comparative analysis of their training and downstream performance. Our efforts culminate in achieving state-of-the-art performance on a challenging materials science benchmark. Furthermore, we investigate the computation and energy efficiency, and propose a computationally efficient method for architecture design. To our knowledge, these pre-trained models represent the largest available for materials science. Our findings provide practical guidance for building LLMs on HPC platforms. 
%Large language models (LLM) have attracted a lot of attention in and beyond AI community. The generative pre-trained Transformer (GPT) is emerged as the dominant LLM architecture, and many variants have since been proposed. However, those variants were pre-trained under different settings, i.e., different input data and/or data preprocessing procedures, different training recipes, etc, and a controlled comparative study is lacking. Here, we carefully study two popular open-sourced GPT architectures, i.e., GPT-NeoX and LLaMA, on the first Exascale supercomputer --- Frontier. With the same text corpus on materials science and end-to-end pipeline, we compare the training and downstream performance and achieve the state-of-the-art performance for one material science benchmark. Furthermore, we investigate the computation and energy efficiency in terms of achievable floating-point operations per second (FLOPS) and FLOPS per Watt. As far as we know, the pre-trained models are the largest for materials science and achieve better performance for domain benchmarks. Furthermore, our findings provide practical guidance for building LLMs on HPC platforms.   
\end{abstract}


\section{Introduction}
Since the inception of the Transformer architecture \cite{transformer}, Transformer-based large language models (LLMs) have emerged as the bedrock upon which numerous AI breakthroughs have been constructed. Between two widely adopted architectures, namely bidirectional encoder representations from Transformers (BERT) \cite{bert} and generative pre-trained Transformer (GPT) \cite{gpt}, it has been shown that the performance of GPT models scales \cite{scalinglaw} with both model and data sizes, while marginal benefit was observed \cite{scholarbert} comparing BERT models of different sizes. Many efforts have then been devoted to improving GPT architectures, including GPT-1 to GPT-4 \cite{gpt}, GPT-NeoX \cite{neox}, LLaMA \cite{llama,llama2}, etc. Although the record performance was refreshed repeatedly, the focus was on downstream tasks only and little was discussed on the reason behind the architecture choices. A careful examination of current practices on the end-to-end pipeline for building LLMs is needed, especially for scientific applications.

With the rise of LLMs, a new research paradigm emerges in the field of AI for sciences, which is to build a foundation model via unsupervised learning on scientific data, and then fine-tune it to apply for many downstream scientific tasks. There have been several early attempts, e.g., the ClimaX \cite{climax} --- a vision Transformer foundation model for climate, bioGPT \cite{biogpt}, and pubmedGPT \cite{pubmedgpt} which are generative pre-trained Transformers based on biology and medical text data. To the best of our knowledge, there is not yet a foundation model pre-trained specifically for materials science. 

Materials science is fundamental to everyday life. From lightweight materials in transportation to innovative energy storage, materials science plays a crucial role in shaping modern society. The design of new materials relies on the understanding of existing research. To extract knowledge from the materials publications, several natural language models have been built, including word2vec models \cite{6m}, BERT-style models \cite{matscibert}, etc. However, these models are limited to specific tasks and cannot be generalized. We intend to pre-train a foundation model based on the GPT architecture and demonstrate its capability on both generic language benchmarks and scientific downstream tasks. 

While the current open-sourced state-of-the-art GPT variant is LLaMA \cite{llama} (more recent  LLaMA2 \cite{llama2} includes tweaks to improve inference performance) and the top performer Falcon model \cite{falcon} employs the same LLaMA architecture but with improved data quality, little work has been done comparing different architectures and pre-training recipes. In this work, we will investigate two prominent open-sourced GPT variants, namely GPT-NeoX \cite{neox} and LLaMA, and show performance comparisons on both model training and downstream tasks. In addition to evaluating zero- and few-shot performance on question answering tasks, we also propose a scientific regression task to showcase the scientific usage of the model. 

Another important aspect in building LLMs is the computational performance and energy efficiency because the process typically incurs a large computational cost and significant energy consumption. Most of the established practices, however, are on NVIDIA GPU-based platforms. Here we will study the performance and optimization of the popular training frameworks on Frontier --- the first Exascale supercomputer which is equipped with AMD GPUs.    
Specifically, our contributions include:     
\begin{itemize}
    \item We comparatively studied two mainstream GPT architecture variants, i.e., GPT-NeoX and LLaMA, and evaluated their end-to-end pre-training recipes.
    \item We pre-trained and openly released a set of foundation models for materials science, called MatGPT. 
    \item We proposed a new downstream task for scientific usage and achieved state-of-the-art performance on a challenging materials science benchmark. 
    \item We established baselines and practical guidance for building LLMs on AMD GPU-based platforms. 
\end{itemize}

{\color{black}Our focus lies in constructing LLMs for scientific purposes on HPC. While we exemplify this within the field of materials science, our methods are not constrained to a singular domain. The rest of the paper is organized as follows: Sec.~\ref{sec:related} reviews the architecture trends of LLMs and current state-of-the-art applications for materials science. Sec.~\ref{sec:method} details our data corpus and end-to-end computational approach, where we introduce a computationally efficient way for exploring the architecture space, and a generic method to integrate LLMs with existing models for scientific applications. Sec.~\ref{sec:evaluation} presents our comprehensive comparisons from architecture selection and training performance, to downstream tasks and model explainability, and we conclude in Sec.~\ref{sec:conclusion}.}

\section{Related work}\label{sec:related}
\begin{figure}[t]
\centering
\includegraphics[width=0.48\textwidth]{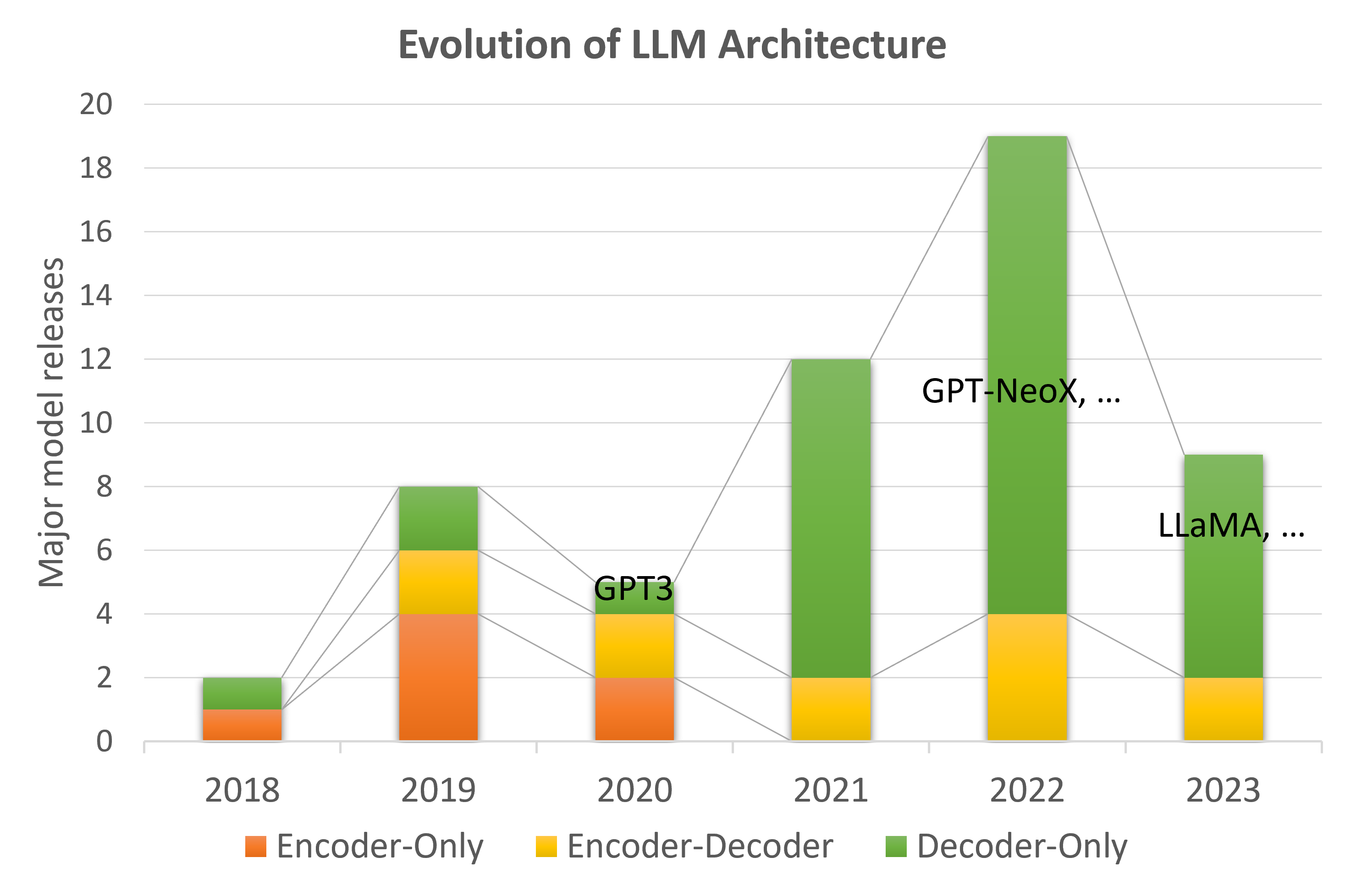}
\caption{Evolution of LLM architecture since 2018. Starting from 2021, the GPT architecture dominates the major model releases. }\label{fig:evol}
\end{figure}
Following the evolutionary tree of LLM architecture \cite{evolution}, three main branches, i.e., encoder-only, encoder-decoder, and decoder-only, stem from the introduction of Transformer architecture \cite{transformer} in 2017. The number of major model releases within each branch are plotted in Fig.~\ref{fig:evol} for each year ever since. From 2018 to 2019, encoder-only models such as BERT \cite{bert} enjoyed more popularity. Since GPT-3 \cite{gpt3}, which demonstrated emerging capability with the unprecedented parameter size (175B), the decoder-only architecture has dominated the field. GPT-NeoX \cite{neox} and LLaMA \cite{llama} are among the most popular open-source variants of GPT-3. On the other hand, the number of encoder-decoder models, e.g, T5 \cite{t5} for translation tasks, has stayed about the same.      

In recent years, the materials science community has embraced the advancements in natural language processing (NLP). A study \cite{nature} in Nature built a word2vec model on 3M abstracts and demonstrated its usage in material recommendation for functional applications. Since then, BERT-style models \cite{matscibert, matbert} pre-trained specifically on material texts have been the best performing models; and MatSciBERT \cite{matscibert} is considered as the current state-of-the-art for domain-specific LLM for materials science. However, the model and data sizes are limited to hundreds of millions of parameters and several millions of papers, respectively. No generalization ability has been demonstrated. 

An early attempt \cite{gpt-finetune} was made to apply GPT-3.5 model to study energy materials, but it is fine-tuned only on a small dataset and a generic foundation model for materials science is still lacking.

\section{Method}\label{sec:method}

\begin{table}[h]
\centering
\caption{Data Sources for MatGPT.} \label{tab:data}
\begin{tabular}{cccc}
\toprule
Source & \#abstract & \#full-text & \#tokens \\ \midrule
CORE   &  2.5M      &    0.3M     &  8.8B       \\
MAG    & 15M        &             &  3.5B       \\
Aminer & 3M         &             &  1.2B       \\
SCOPUS & 6M         &             &  1.5B       \\ \midrule
All    &  26.5M     &    0.3M     & 15B         \\ \bottomrule
\end{tabular}
\end{table}

\noindent \textbf{Data Sources} We collect the abstracts and full-text data from the four data sources, including CORE, Microsoft Academic Graph (MAG), Aminer, and SCOPUS, as listed in Table~\ref{tab:data}. For SCOPUS, we use the publisher's API to retrieve the abstracts of about 6M materials science publications \cite{6m}. For the other sources, aggregated data covering all scientific domains are downloaded and then preprocessed to filter out materials science-related ones. The screening is performed via a fine-tuned SciBERT model on a small domain-labeled dataset, and the resulting classifier can then be used to partition the aggregated data sources. In total, there are 26.5M abstracts and 0.3M full-texts, counting to about 15B tokens.         

\begin{table*}[t]
\centering
\caption{Model architectures and various data tokenization ( HuggingFace and Sentencepiece tokenizer --- HF and SPM, and vocabulary size of 32K and 52K). } \label{tab:arch}
\small
\begin{tabular}{cccccccc}
\toprule 
MatGPT Arch              &  \#parameters       & hidden-size & \#layers & \#heads & head-dim  & tokenizer & vocab-size \\ \midrule
\multirow{2}{*}{LLaMA}    & 1.7B & 2304        & 24       & 24 &  96     & SPM/HF    & 32K/52K    \\
                          & 6.7B & 4096        & 32       & 32    & 128  & HF        & 52K        \\ \midrule
\multirow{2}{*}{GPT-NeoX} & 1.7B & 2304        & 24       & 24     & 96 & HF        & 52K        \\
                          & 6.7B & 4096        & 32       & 32   & 128   & HF        & 52K        \\ \bottomrule 
\end{tabular}
\end{table*}

\noindent \textbf{Model Architecture}
\begin{figure}[th]
\centering
\includegraphics[width=0.48\textwidth]{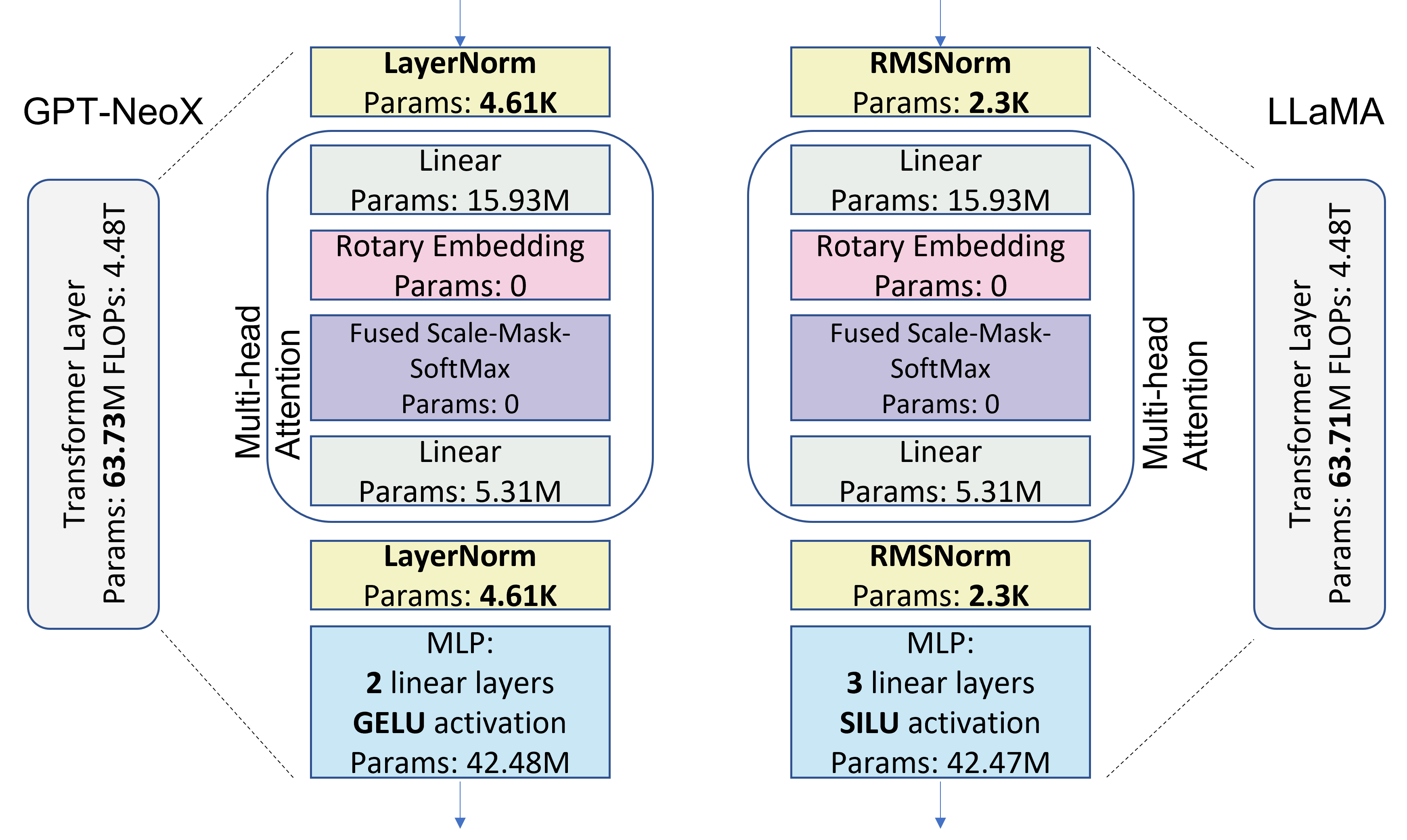}
\caption{Transformer layer of GPT-NeoX and LLaMA architecture, respectively. The specific parameter and FLOP numbers are for 1.7B parameter model with a sequence length of 2048 and batch size of 16.}\label{fig:arch}
\end{figure}
We build MatGPT upon two mainstream model architectures, GPT-NeoX \cite{neox} and LLaMA \cite{llama}, both of which are based on GPT-3 but with different variations. As illustrated in Fig.~\ref{fig:arch}, the original LLaMA employs the SentencePiece (SPM) tokenizer with a vocabulary size (vocab-size) of 32K; While GPT-NeoX utilizes the HuggingFace (HF) tokenizer with a vocab-size of 52K. SPM has fine-grained control over subword tokenization while HF is more popular for pre-trained Transformers. Both architectures use rotary positional embeddings \cite{su2021roformer} instead of absolute positional embeddings \cite{gpt1} as in GPT-3, and LLaMA made further modifications to the pre-normalization and activation by using RMSNorm and SwiGLU \cite{llama} activation functions. The specific choice of number of layers and hidden size, as listed in Table~\ref{tab:arch},  also takes the computational performance into consideration, which will be discussed in Sec.~\ref{sec:results}. For the models of the same specification (i.e, number of layers and attention heads, and hidden size), each Transformer layer of GPT-NeoX and LLaMA has approximately the same number of parameters and floating-point operations (FLOPs). The multi-head attention layers are exactly identical, and the main difference is the normalization layers and multi-layer perceptrons (MLPs).  

The Chinchilla study \cite{20x} suggested that the optimal token-to-parameter ratio is 20, while more recent studies \cite{llama, llama2} indicates a higher ratio, i.e., a smaller model and larger data, can be beneficial. Given our data size (15B tokens), we build MatGPT with a few billion parameters. To identify the most computationally efficient architecture, we perform a grid search for various numbers of layers and hidden sizes. The specific choice has to satisfy following constraints, 
\begin{align}
    N_{h} \ \% \ N_{a} &= 0 \\
    N_{h} \ \% \ (TP) &= 0 \\ 
    N_{l} \ \% \ (PP) &= 0 \\
    N_{a} \ \% \ (TP) &= 0  \\
    (TP\times PP\times DP) \ \% \ 8 &= 0      
\end{align}
where $N_{h}$, $N_{l}$, and $N_{a}$ denote hidden size, number of layers and attention heads, and TP, PP, and DP are the tensor, pipeline, and data parallelisms, the product of which should be equal to the total number of devices (multiple of 8 for our platform). The dimension of the attention head is implemented as the ratio of $N_{h}$ over $N_{a}$, hence the constraint; and the rest is to ensure the workload can be evenly distributed among parallelism dimensions.       

We will compare the model performance of the two architectures, as well as different choices of tokenizers and vocab-sizes. As far as we know, this is the first controlled study of LLM architectures at a large scale.          

\noindent \textbf{Flash Attention} One of the most important recent advances in the optimization of training Transformers is flash attention \cite{flash}, which significantly reduce high-bandwidth memory (HBM) accesses which in turn speeds up the computation of the attention calculations. Because it leverage classical HPC techniques at a lower level (e.g., tiling, recomputation), the implementation is hardware-specific. In our evaluation, we make use of the open-sourced development \cite{flash-rocm} of flash attention on AMD GPUs. It is based on the composable kernel library \cite{ck}, which performs many of the same functionalities of the CUTLASS library for NVIDIA GPUs, and currently both forward and backward calculations are supported. {\color{black}The latest flash attention v2\cite{flash2} is also in the process of being incorporated into AMD GPUs. We will study the computational and memory impact of both versions.} To our best knowledge, this is the first such study on recent AMD hardware (MI250X). 

\noindent \textbf{Training and Evaluation Framework}
Our evaluation is based on the GPT-NeoX \cite{neox} framework, a DeepSpeed-Megatron \cite{deepspeed,megatron} implementation. We port it to the AMD GPU-based platform, Frontier \cite{frontier} with following modifications: 1) Add support for the LAMB optimizer \cite{lamb}, which is an enhanced Adam optimizer \cite{adam} that can mitigate the generalization gap caused by the large-batch training. 2) Hipify the fused kernel in CUDA to support the AMD ROCm stack. 3) Adapt the flash attention interface from the corresponding AMD implementation.   

Because the LLM training is shown to be communication-bound on Frontier \cite{cost} and model parallelism is more demanding than data parallelism in terms of the communication requirement \cite{anthony2023mcrdl}, it is desired to assign most compute resources (i.e., GPUs) to data parallelism (i.e., large batch size), in order to achieve good scaling efficiency and energy utilization. Therefore, exploring the LAMB optimizer is necessary for efficient distributed training of LLMs on HPC systems.
   
Regarding the downstream evaluation, we employ the generic language model evaluation framework \cite{eval-harness} for the common question answering tasks. This will demonstrate MatGPT's generalizability as a foundation model, even though it is not pre-trained on generic web texts. More importantly, we propose a new scientific downstream task to demonstrate the performance and science usage of MatGPT, namely the material properties prediction. 

\iffalse
\noindent \textbf{Scientific Downstream Task --- Classification}
There are 230 crystal space groups in nature, and all existing solid-state materials fall under one of the space-group classes. This is nature's ImageNet classification for materials science. We use the dataset \cite{cbed} collected from the materials project \cite{mat-proj}, which consists of over 60K materials. We then fine-tune MatGPT on the materials' formula (e.g., Li4Co2C4O12) for the classification of the space groups. Some materials can have different symmetries (i.e., different space groups) depending on external conditions (e.g., temperature, pressure, etc). For simplicity, we de-duplicate the label and keep only one space group for each material.   
\fi 

\noindent \textbf{Scientific Downstream Task}
%Guojing Cong 
{\color{black}%A straightforward and generic method to integrate LLMs with scientific applications is augmenting the feature space of existing models with LLM embeddings. Because the correlation of specific terms, e.g., named entities, are captured by LLM embeddings, the numerical vector of text terms in the LLM latent space can be used as features to predict the characteristics associated with the corresponding terms. For LLMs pre-trained on scientific publications, such as MatGPT, the embeddings should outperform those of LLMs pre-trained on web texts.  
A direct and versatile approach for integrating LLMs into scientific applications involves enhancing the feature space of existing models by incorporating LLM embeddings \cite{forge}. These embeddings effectively capture correlations among specific terms, such as named entities, within the LLM context. Consequently, the numerical vector representation of text terms within the latent space of LLMs can serve as features to predict characteristics associated with corresponding terms. For LLMs pre-trained on scientific publications, such as MatGPT, the embeddings are expected to outperform those of LLMs pre-trained on web texts, and the larger the capacity (i.e., model parameters) of a LLM, the better quality of its embeddings. To demonstrate the capability of MatGPT, we will show the comparisons of its embeddings with other LLMs including MatSciBERT.  

As an example for materials science, }we explore leveraging the trained LLMs for improving learning in material properties prediction. Graph neural networks (GNNs) are the de facto paradigm for predicting the physical and functional properties of materials. Recent studies incorporate features of increasing complexity such as Gaussian radial functions, plane wave functions, and angular terms to augment the neural network models, with the expectation that these features are critical for achieving high performance~\cite{xie2018crystal,rosen2021machine,park2020developing,choudhary2021atomistic,cheng2021geometric,wang2020accelerating}. However, none of these efforts tap into the vast amount of publications that reflect human understandings of material research. With LLM trained on material texts, we experiment with using both material structure data, fed to GNNs, and scientific publications, reflected in the embeddings of corresponding material formulas, for learning.

 \begin{figure}[t]
\centering
\includegraphics[width=0.5\textwidth]{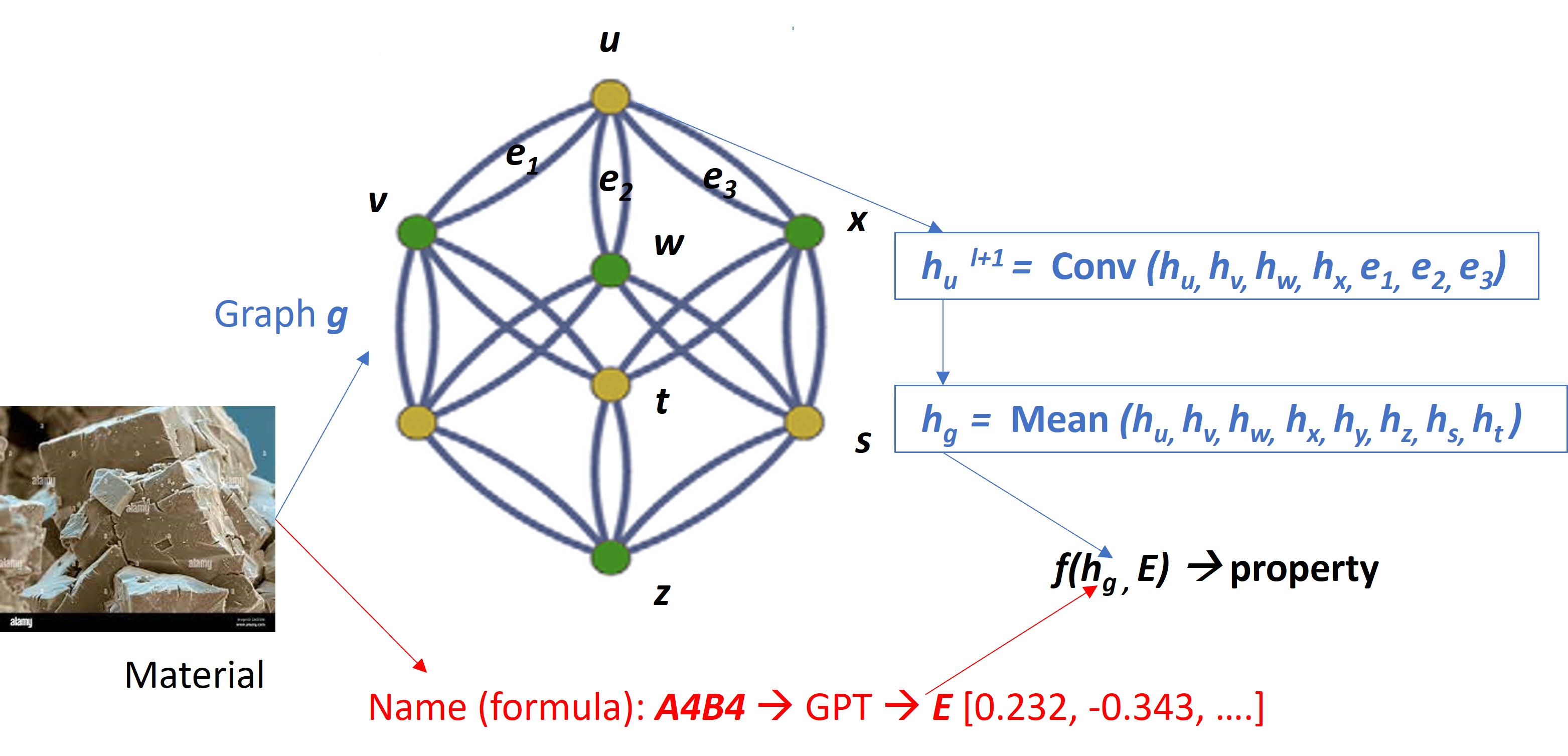}
\caption{A new scientific usage of LLM: combining LLM embeddings with GNN for material properties prediction.}\label{fig:gnnlmm}
\end{figure}

\begin{figure*} [h]
\centering
\includegraphics[width=0.45\textwidth]{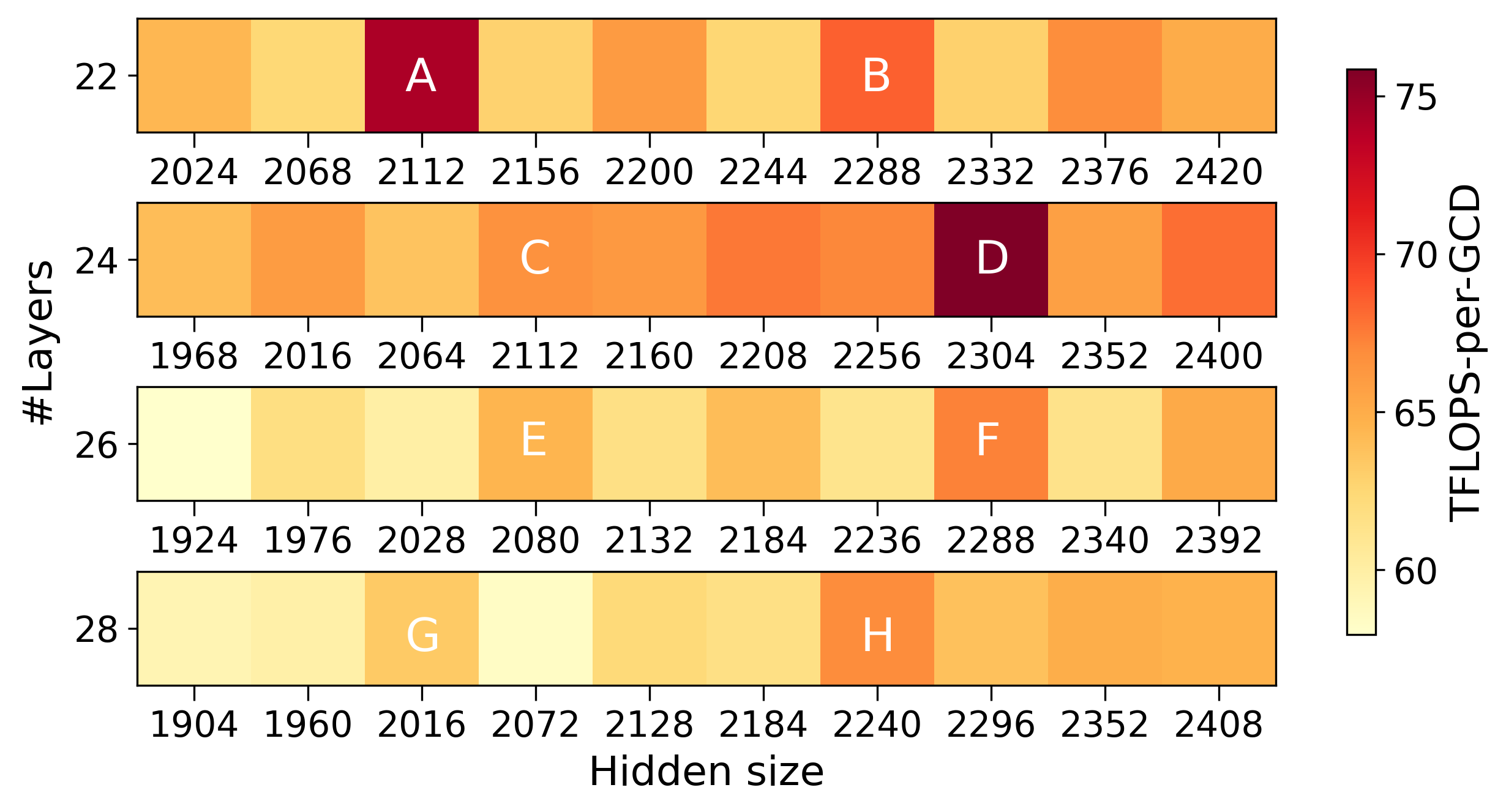}
\includegraphics[width=0.45\textwidth]{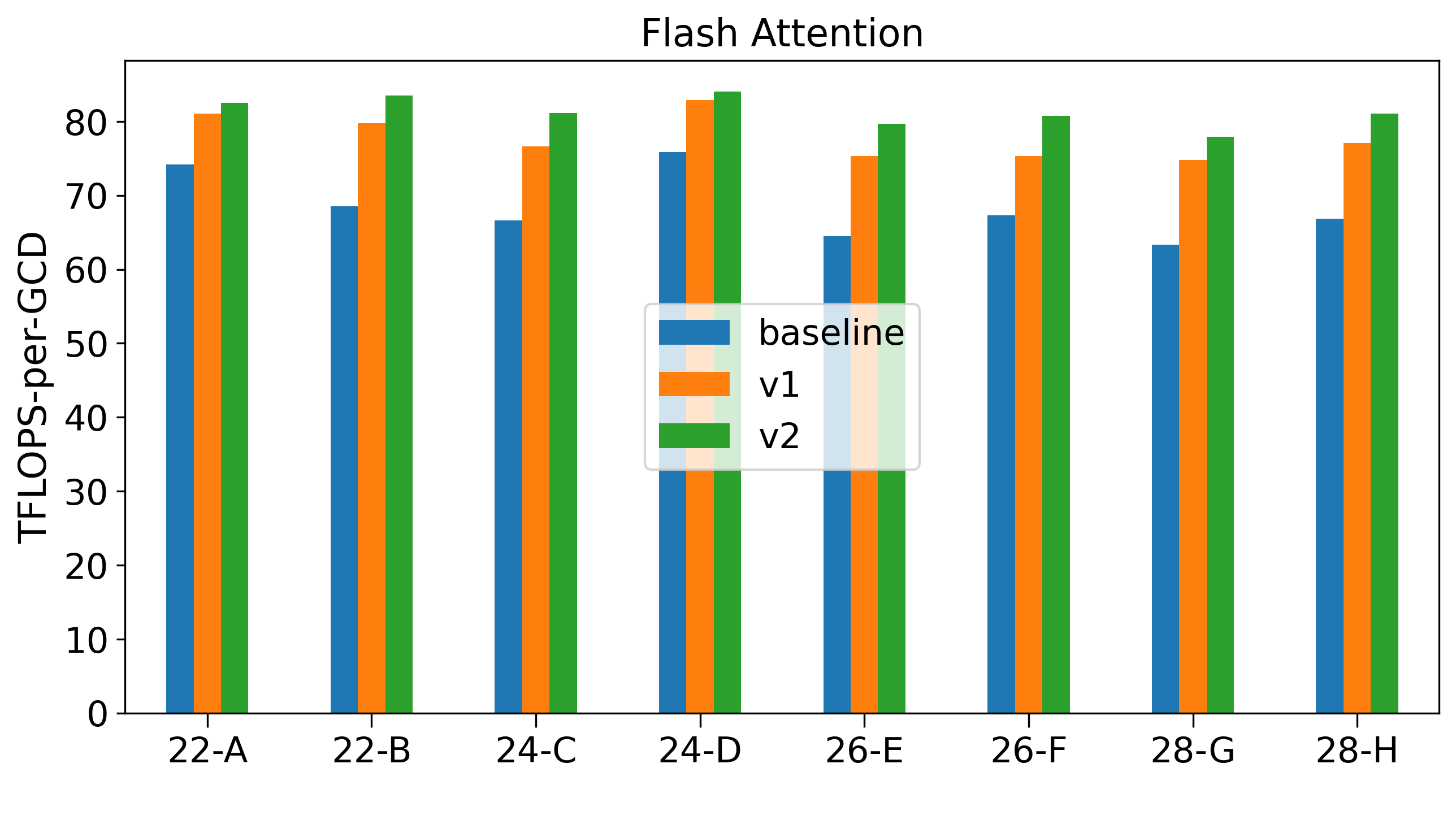}
\caption{(Left) The heatmap of training throughput (TFLOPS per GPU) for MatGPT with various numbers of layers and hidden sizes for model size around 1B. {\color{black}(Right) The performance boost for architectures eligible for flash attention, including v1 and v2, respectively.}}\label{fig:heatmap}
\end{figure*}

\begin{figure}[h]
\centering
\includegraphics[width=0.45\textwidth]{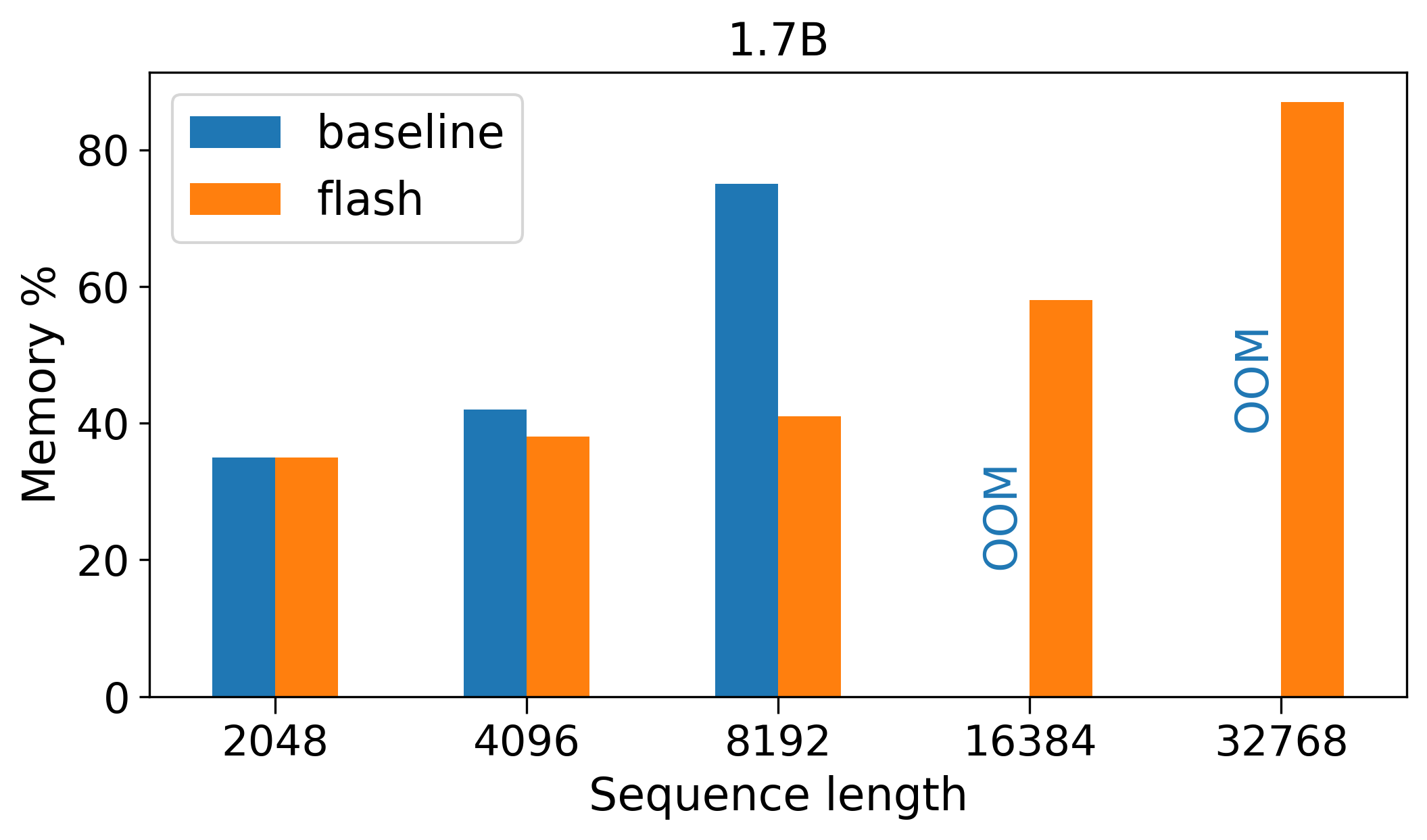}
\caption{\color{black}The peak memory usage (percentage) during the training of MatGPT 1.7B with and without flash attention for context sequence length from 2,048 to 32,768.}\label{fig:long}
\end{figure}

 For demonstration, we choose to predict band gap, a quantum chemical property of materials. The ground truth for the band gap is computed from density function theory (DFT) calculations. DFT is a quantum mechanical method for solving the electronic structure of a material widely used in computational materials chemistry. The band gap of a material is the energy difference between its highest occupied electronic energy level and the lowest unoccupied energy level. It is more challenging to predict band gap than other properties such as formation energy. 

 We adopt a simple methodology that concatenates the output from a GNN for a material with the LLM embedding of the corresponding material's chemistry formula. The learning paradigm is shown in Figure~\ref{fig:gnnlmm}.

In Figure~\ref{fig:gnnlmm}, a graph $g$ is constructed from the structure information (e.g., atoms, their 3-dimensional positions, and their interactions) of the input material, and graph convolution is then applied to derive a feature $h_g$. At the same time, an embedding $E$ is generated using the trained LLM for the formula of the material. $E$ and $h_g$ are then concatenated as a feature for the prediction of the property.

%Molecular formation of materials demonstrated as graph structure leads to effective extraction of both structural and internal feature information from consisting atoms of a given molecule using graph neural networks (GNNs). Although GNNs are utterly useful for complex regression tasks such as quantum chemical property prediction (e.g. band gap, formation energy, etc.), limited GNN model performance due to a lack of semantic and contextual information is a persisting challenge which can be addressed by leveraging knowledge from language models. The test data set for the regression task encompassing a small domain (metal-organic data set) lacks discriminative features between materials in latent space which is crucial for enhancing the model's performance. 

Our study demonstrates the boost of learning performance of GNNs from incorporating additional semantic and contextual information.  Moreover, the results and insight from the experiments by employing MatSciBERT and MatGPT variants robustly support our claim of having more expressive model architectures, MatGPT trained on larger datasets with a large number of parameters compared to MatSciBERT.
The dataset we use is the material project dataset~\cite{jainmaterials}. 

\iffalse
\begin{enumerate}
    \item gpt-neox
    \item llama
    \item table hyper-parameters
    \item table for forge-mat arch
    \item scientific usage: classification, regression
\end{enumerate}
\fi

\section{Evaluation}\label{sec:evaluation}

\subsection{Experiment setup}
We perform the experiments on the first Exascale supercomputer, Frontier. Each Frontier node is equipped with four AMD Instinct MI250X GPUs with dual Graphics Compute Dies (GCDs) and one third-generation EPYC CPU. A GCD is viewed as an effective GPU, and we use GCD and GPU interchangeably in the following discussion. All four MI250Xs (eight effective GPUs) are connected using 100 GB/s Infinity Fabric (200 GB/s between 2 GCDs of MI250X), and the nodes are connected via a Slingshot-11 interconnect with 100 GB/s of bandwidth. Frontier consists of 9408 nodes in total, i.e., 75,264 effective GPUs (each equipped with 64GB high-bandwidth memory). 

{Our evaluation is based on GPT-NeoX \cite{neox} implementation with PyTorch v1.14.0 \cite{pytorch} and DeepSpeed v0.7.3 \cite{deepspeed}, which are built against AMD ROCm v5.4.} In the following, we will investigate the MatGPT performance, including training throughput, architecture comparisons, and model performance on our newly proposed scientific downstream task and commonly used language benchmarks. %{\color{black}The tokenized data and pre-training code are publicly available\footnote{\color{black}https://github.com/xxx, anonymized for review.}.} 

% \usepackage{multirow}
\begin{table}[h]
\centering
\caption{Training hyper-parameters for MatGPT.}\label{tab:params}
\begin{tabular}{llllll}
\hline
Model                 & Optimizer & $\beta_1$ & $\beta_2$ & LR     & BS \\ \hline
\multirow{2}{*}{1.7B} & Adam      & 0.9   & 0.95  & 0.0002 & 1M \\
                      & LAMB      & 0.9   & 0.999 & 0.01   & 4M \\ \hline
6.7B                  & LAMB      & 0.9   & 0.999 & 0.006  & 4M \\ \hline
\end{tabular}
\end{table}
% experiment parameters 

Considering the large GPU capacity (both memory and count) and network bandwidth (relatively limited compared to AI-oriented machines such as Selene\cite{megatron}) on Frontier, it is desirable to leverage large-batch training in order to achieve good scaling efficiency and reduce time-to-solution. We train MatGPT with the LAMB optimizer \cite{lamb}, a variant of the Adam optimizer with the layer-wise learning rate adjustment to mitigate the generalization gap of large-batch training. We use $\beta_1=0.9, \beta_2=0.999$, a weight decay of 0.1, and a batch size of 4M tokens. The cosine learning rate scheduler is employed with an initial learning rate 0.01 and a final learning rate set to 10\% of the initial learning rate. We use 1\% of the total batch steps for warmup. {\color{black}The training is performed in bfloat16, which provides better numerical stability.}

\subsection{Results} \label{sec:results}
Following the above experiment setup, we conduct a thorough investigation in the end-to-end pipeline of building MatGPT, including the evaluation of the computational performance for training, comparisons of both the loss and zero-shot tasks performance for different architecture choices, and fine-tuning performance for the scientific downstream task we proposed.    

\noindent \textbf{Training Performance}
The scaling law \cite{scalinglaw} indicates the model performance scales with the number of parameters, however, for similar-size models, the choice of the number of layers, hidden size, and attention heads, seems arbitrary. We refine our model design by taking the computational performance into consideration. Because the workhorse of the training is matrix multiplication and the underlying math library, such as cuDNN for NVIDIA GPUs and MIOpen for AMD GPUs, is optimized for certain shapes of matrices \cite{spock}. 

\begin{figure}[t]
\centering
\includegraphics[width=0.45\textwidth]{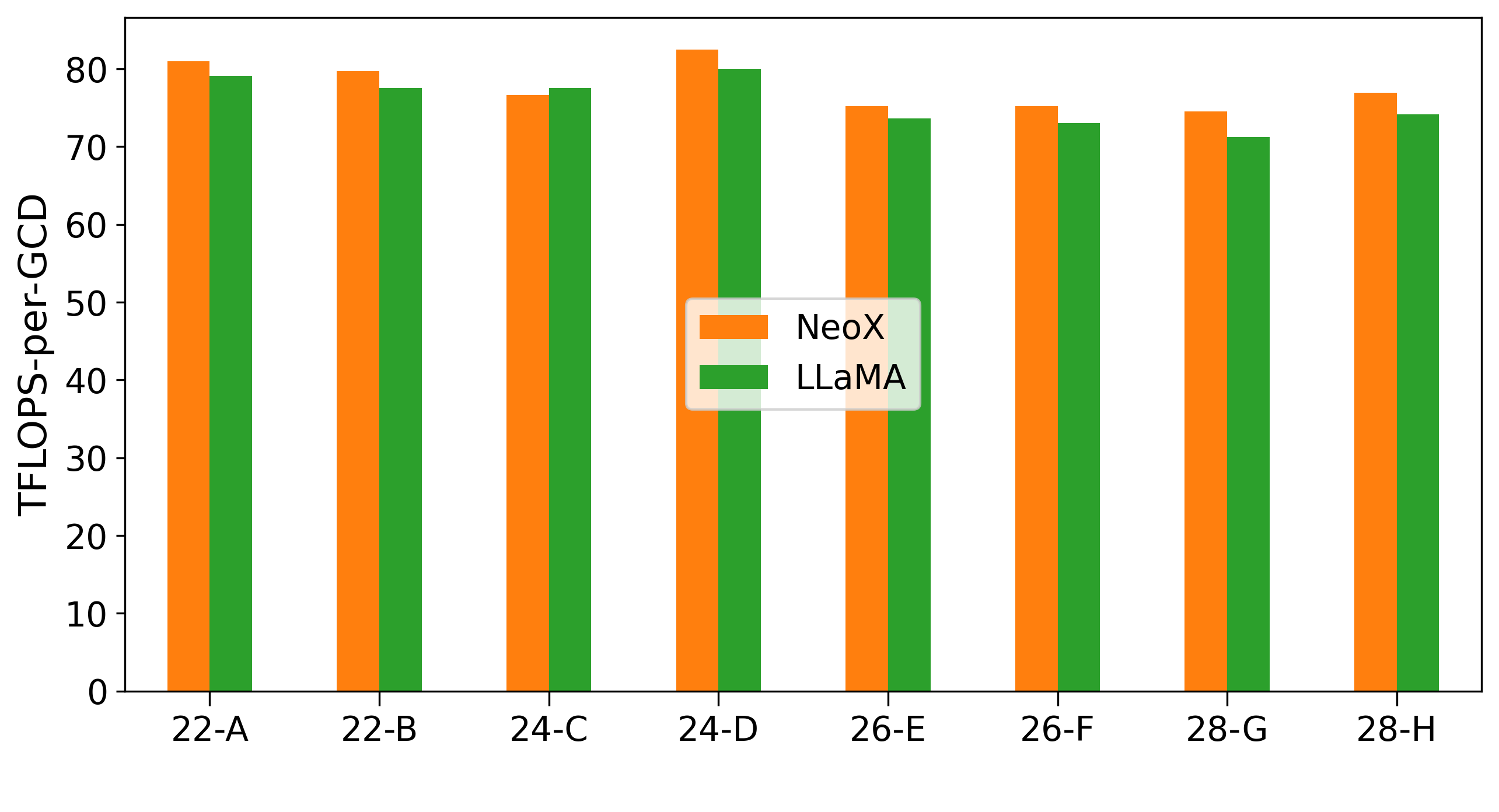}
\caption{The comparisons of training throughput (TFLOPS per GCD) for MatGPT -NeoX and -LLaMA achitectures.}\label{fig:comp}
\end{figure}

In Fig.~\ref{fig:heatmap}, we plot the heatmap of MatGPT-NeoX  training throughput in terms of TFLOPS per GPU for various numbers of layers and hidden sizes for model sizes around 1B parameters. The performance varies from 58 to 76 TFLOPS, and the best case (achieved about 40\% of the MI250X theoretical peak --- 383 TFLOPS for 2 GCDs) corresponds to 24 layers with a hidden size of 2304. This is mostly due to the fact that the computation of multi-head attentions (see Fig.~\ref{fig:arch}) is expensive and a head dimension of multiples of 8 is computationally favorable (it can take full advantage of matrix cores on AMD GPUs, which are equivalent to tensor cores on NVIDIA GPUs). We marked all the architectures (from A to H in Fig.~\ref{fig:heatmap}) with head dimensions satisfying this criteria, and indeed they are among top performers for each layer size. Coincidentally, flash attention also requires the head dimension to be multiples of 8 (and {\color{black}up to 256 for v2}), and the performance boost for each case are also shown in Fig.~\ref{fig:heatmap} {\color{black}for v1 and v2, respectively}. On average, the current flash attention implementation improves the computational performance by about 14\% {\color{black}(v1) and 19\%(v2)}, with the best overall training throughput of about 82 TFLOPS per GCD (164 TFLOPS per MI250X) {\color{black}for v1 and 84/168 TFLOPS for v2}, respectively, on Frontier. Note that {\color{black}these numbers are averaged per-node performance for a sequence length of 2048 and} the port \cite{flash-rocm} of flash attention to the ROCm stack is still in active development, and further improvement is expected.   

{\color{black}Furthermore, the flash attention reduces the memory complexity from quadratic to linear in terms of the sequence length, and hence enables longer context window. In Fig.~\ref{fig:long}, we plot the peak memory usage during the training of MatGPT 1.7B for context sequence lengths ranging from 2048 to 32768. Without flash attention, the training process runs out of memory (OOM) for sequences longer than 8192; With flash attention enabled, the memory growth becomes linear (after the sequence length dominates the memory usage) and the maximum supported sequence length increases by about 4X to 32768 on Frontier.} 

To compare the computational performance for NeoX and LLaMA, we plot the training throughput for all 8 cases with flash attention, as shown in Fig.~\ref{fig:comp}. Because of the architecture similarity, especially the identical attention layer (see Fig.~\ref{fig:arch}), both perform more or less the same, with NeoX showing a slight edge in 7 out of 8 cases. The difference likely comes from the parameterization of MLP layers (2 linear layers with GELU activation versus 3 linear layers with SILU activation).

\noindent {\bf Observation \circled{1}} It is computationally desirable to design the LLM architecture with the dimension of attention head to be multiples of 8. With flash attention, the achievable computational performance for training Transformers is over 43\% of the theoretical peak on MI250X {\color{black}for a sequence length of 2048. Given a targeted model size, computational efficiency can be a criterion for architecture selection.}   

\begin{figure}[t]
\centering
\includegraphics[width=0.45\textwidth]{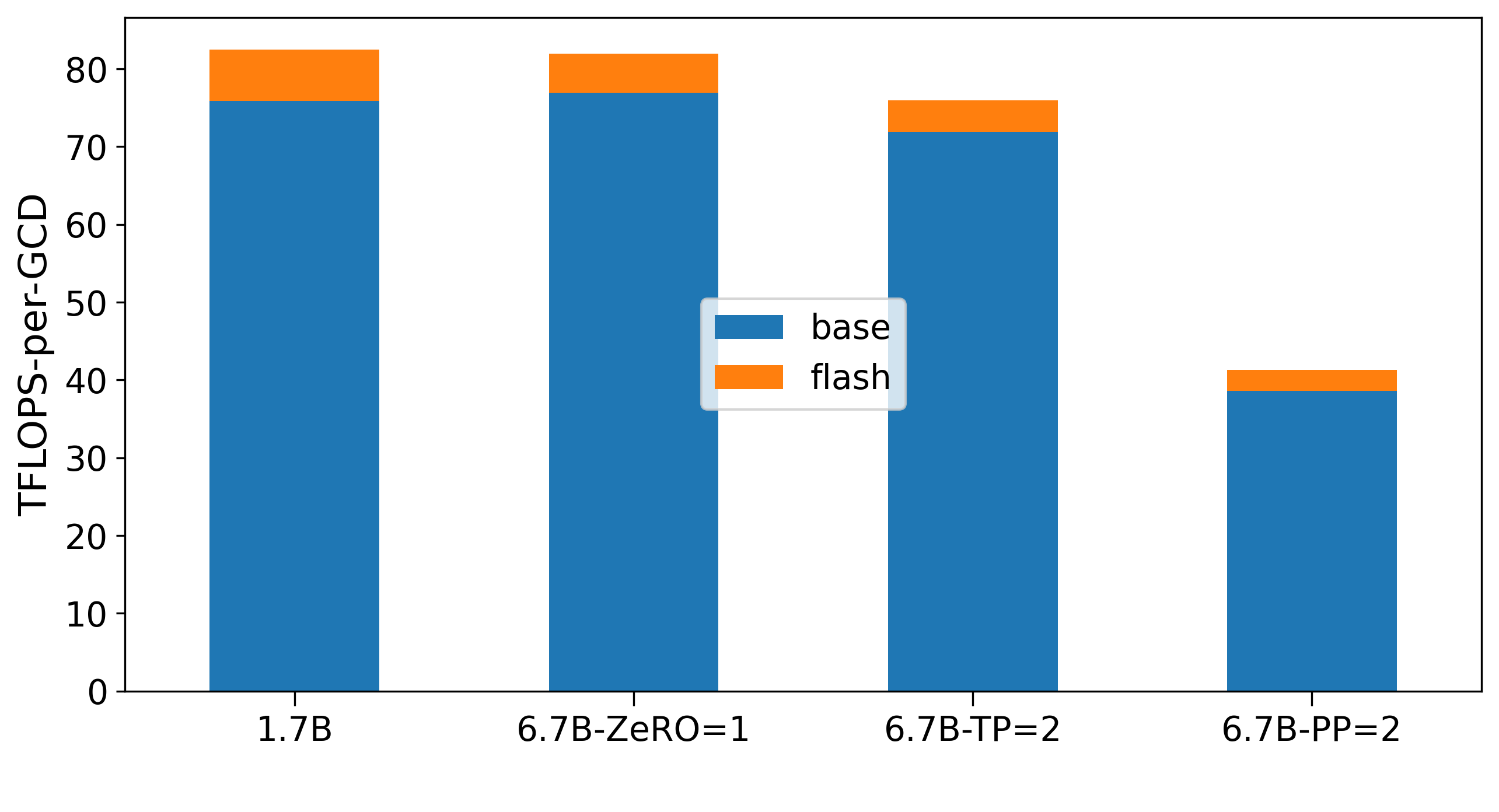}
\caption{The training throughput (TFLOPS per GCD) for MatGPT 1.7B and 6.7B with different parallelisms: ZeRO stage 1 (ZeRO=1), tenor parallelism of 2 partitions (TP=2), and pipeline parallelism of 2 stages (PP=2). }\label{fig:comp2}
\end{figure}

\begin{figure}[h]
\centering
\includegraphics[width=0.45\textwidth]{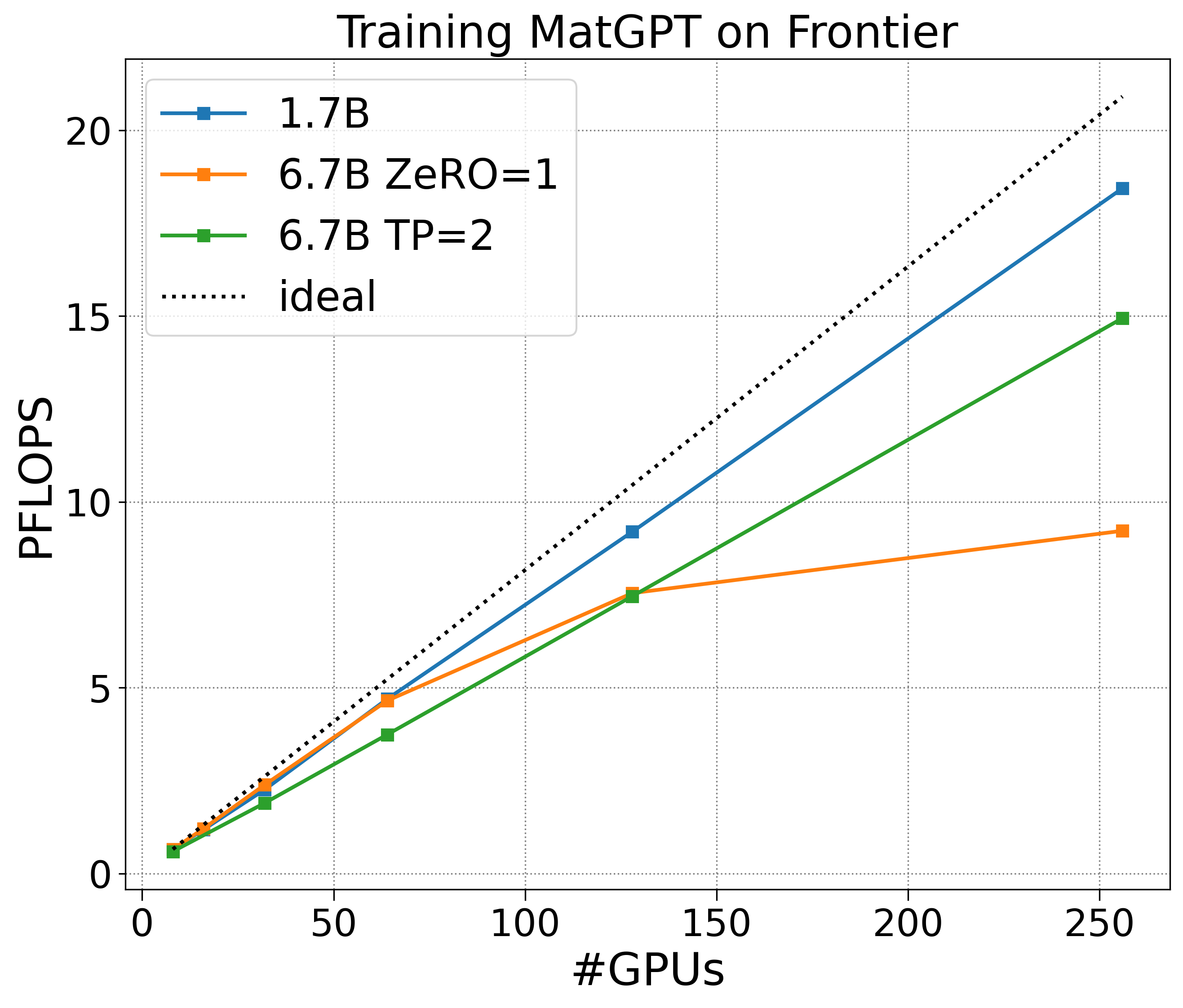}
\includegraphics[width=0.48\textwidth]{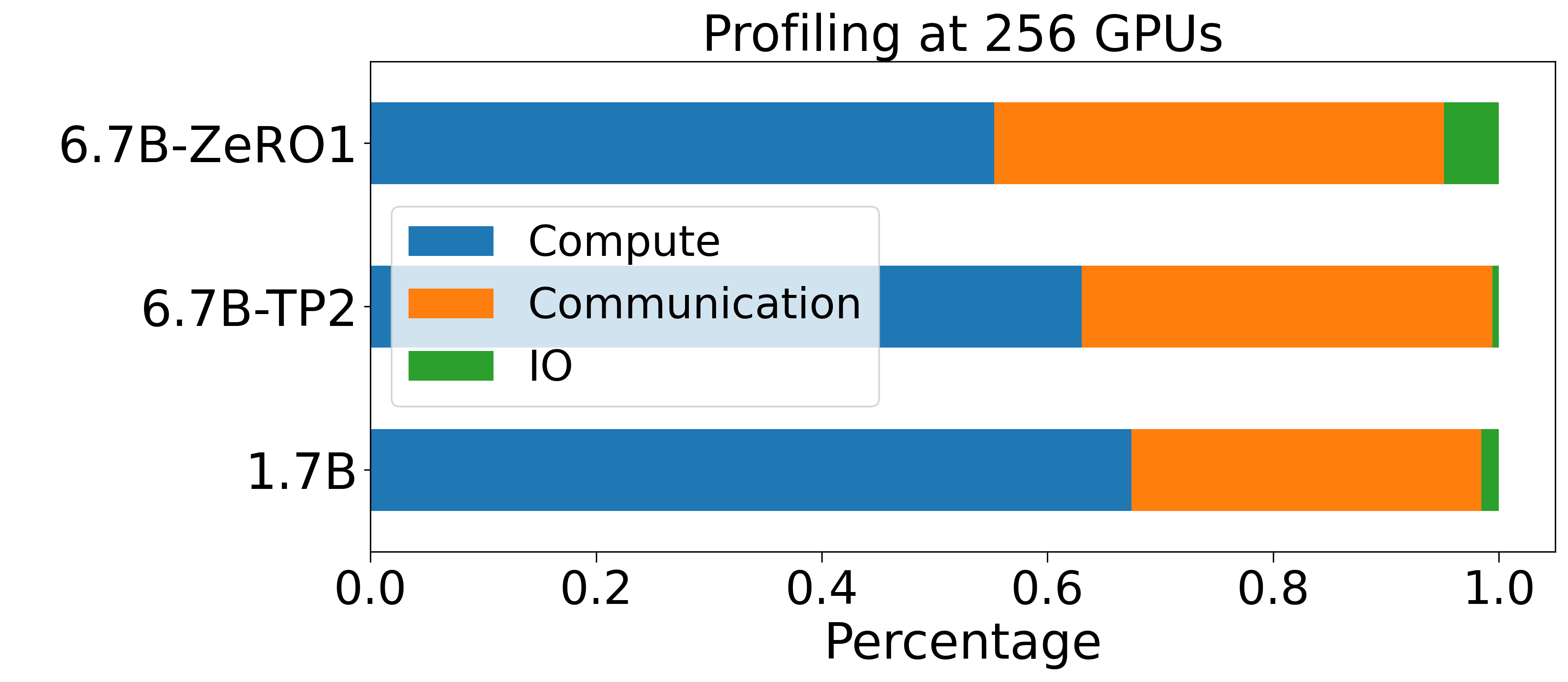}
\caption{(Top) The Scaling of training throughput (PFLOPS) for MatGPT 1.7B and 6.7B with different parallelisms: ZeRO stage 1 (ZeRO=1), tenor parallelism of 2 partitions (TP=2). {\color{black}(Bottom) The profiling breakdown of compute, communication, and IO for the three corresponding parallel distributions with 256 GPUs.}}\label{fig:scale}
\end{figure}

\begin{figure*}[h]
\centering
\includegraphics[width=0.9\textwidth]{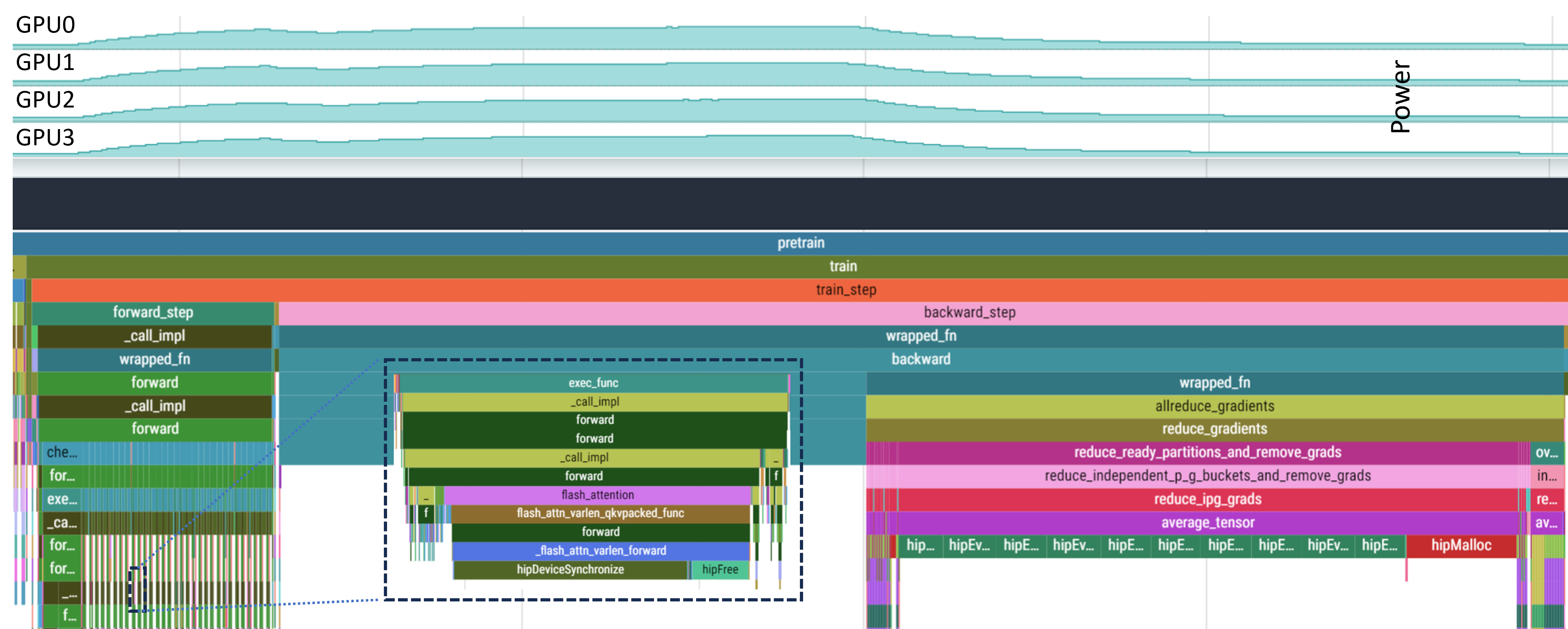}
\caption{\color{black}The runtime and GPU power traces (single node) of distributed training of MatGPT 6.7B with ZerO stage 1 using 256 GPUs. The boxed snapshot is the zoom-in of the forward operations for one of 32 layers.} \label{fig:profile}
\end{figure*}

\begin{figure*} [h]
\centering
\includegraphics[width=0.45\textwidth]{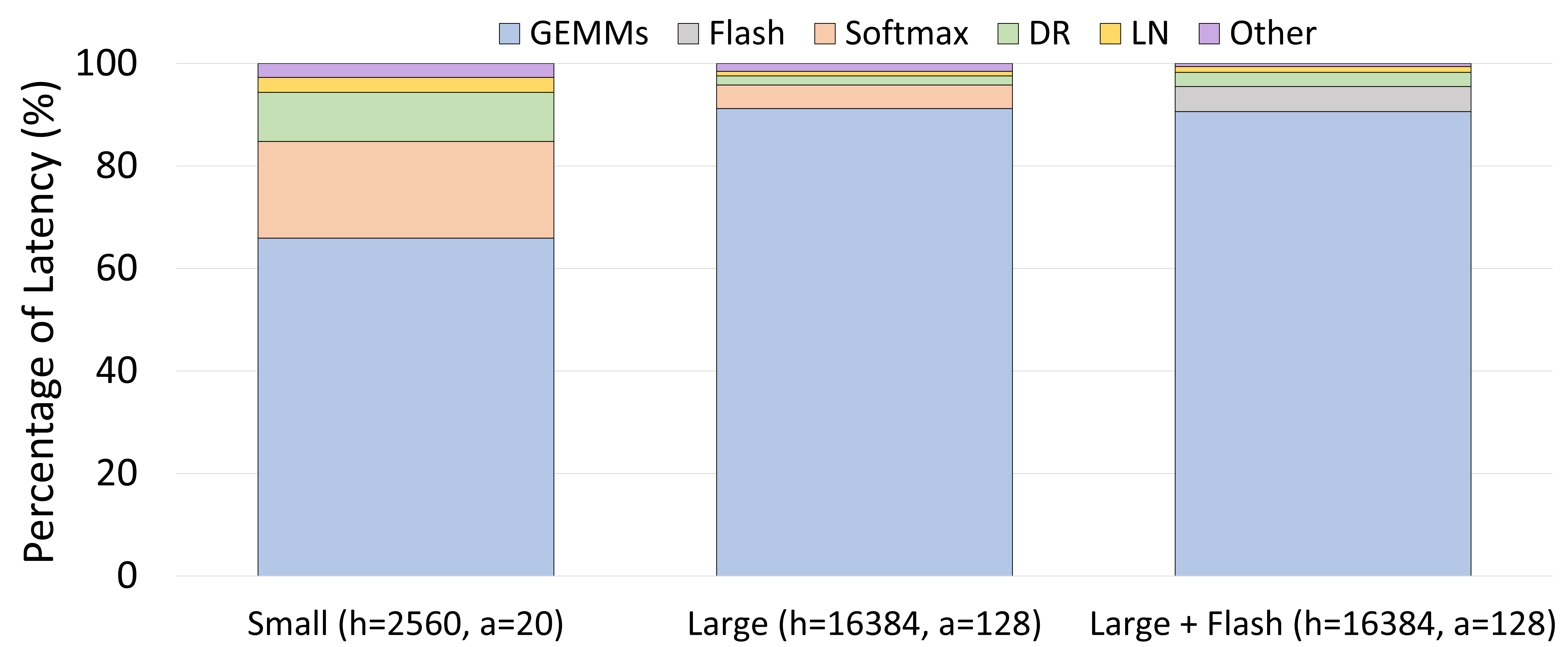}
\includegraphics[width=0.45\textwidth]{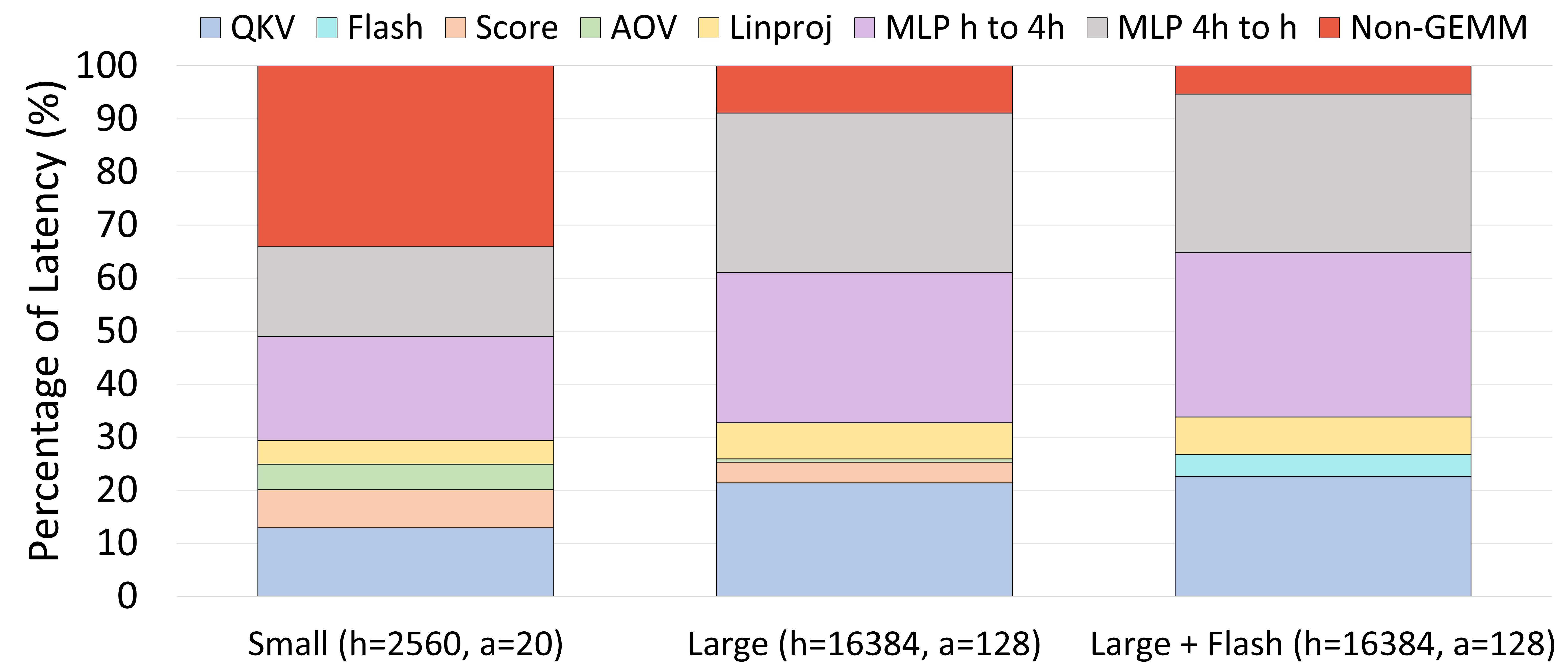}
\caption{{\color{black}(Left) The proportion of latency from each transformer component for one transformer layer of hidden dimension $h$ and number of attention heads $a$. DR and LN stands for dropout and layer normalization operation, respectively. (Right) The individual GEMM proportions of latency for one transformer layer, including query-key-value (QKV), flash attention (flash), attention score (score), attention over value (AOV), linear projection (Linproj), multi-layer perceptron (MLP). }}\label{fig:kernel-breakdown}
\end{figure*}

\iffalse
\begin{figure*}
\centering
%\includegraphics[width=0.45\textwidth]{figures/loss-llama.png}
%\includegraphics[width=0.45\textwidth]{figures/loss-val-llama.png}
\includegraphics[width=0.45\textwidth]{figures/loss-compare.png}
\includegraphics[width=0.45\textwidth]{figures/loss-val-compare.png}
\caption{The training and validation losses of MatGPT- NeoX and LLaMA models. Each curve corresponds to a pre-training experiment with the configuration of architecture-model size-optimizer-batch size. Specifically, architecture: NeoX and LLaMA; model size: 1.7B and 6.7B; optimizer: Adam and LAMB; batch size: 1M and 4M.}\label{fig:loss2}
\end{figure*}
\fi

After identifying the most computationally efficient architecture (i.e., 24 layers and a hidden size of 2304) on MI250X for models of around 1B parameters, we extrapolate the observation to identify a 6.7B model (see Table~\ref{tab:arch}) with a head dimension of 128. As a rule of thumb, the memory footprint for training a GPT-style model is roughly 12 times of the parameters \cite{cost}. For the training of a 1.7B model, a single GCD on a MI250X (equipped with 64GB high-bandwidth memory) is able to accommodate the entire model. However, for a 6.7B model, some level of model parallelism is required. The choices can be DeepSpeed ZeRO optimization (e.g., stage 1 for partitioning the optimizer states), tensor parallelism (e.g., TP = 2 for partitioning each layer onto 2 devices), or pipeline parallelism (e.g., PP = 2 for executing layers through 2 stages, each stage on a separate device). Depending on the parallelism, the communication frequency and message size are different, i.e., imposing different requirements on the platform. In addition to the communication cost of the data parallelisms, the ZeRO parallelism fully shards the optimizer states (twice the memory footprint of model parameters for the Adam and LAMB optimizers) and hence requires all-devices collective communication during the backward propagation. On the other hand, the tensor and pipeline parallelisms have fine-grained control and can limit the extra communication for model parallelism within a subgroup of devices. Although tensor parallelism incurs more frequent messaging than pipeline parallelism (per-layer versus group of layers), there are sequential stages (leading to the so-called ``bubble") in pipelining, and tensor parallelism can perform better with adequate network bandwidth.

In Fig.~\ref{fig:comp2}, we show the performance of training the 6.7B model on a single Frontier node, and compared with that of the 1.7B model. Compared to tensor and pipeline parallelism, ZeRO stage 1 provides the best training throughput (81 TFLOPS per GPU), with a similar boost from flash attention as for the 1.7B model due to the lesser communication frequency.

With the single-node performance optimized, we scale up the distributed training to 256 GPUs and explore 3D parallelism on Frontier.  As shown in Fig.~\ref{fig:scale}, for training MatGPT 1.7B model with data parallelism only, the aggregated performance of 256 GPUs on Frontier can achieve over 18 PFLOPS with a scaling efficiency of 88\%. In comparison, for MatGPT 6.7B model, the per-device throughput is about the same for 64 or less GPUs with ZeRO stage 1 parallelism, and starts to drop at larger scale because of the extra communication overhead of all-device collectives. Tensor parallelism with a partition level of 2, on the other hand, can sustain a 71\% scaling efficiency, owing to the fact that the 2 GPUs/GCDs are within the same MI250X with twice the network bandwidth (200 GB/s). Pipeline parallelism with 2 stages performs much worse compared to the other two parallelism dimensions even for a single node (see Fig.~\ref{fig:comp2}), and hence is not studied at scale. {\color{black}To better understand the scaling behavior, in Fig.~\ref{fig:scale}, we show the profiling for the 3 corresponding parallel distributions (i.e., data parallel for 1.7B, and ZeRO stage 1 and tensor parallel TP=2 for 6.7B) with 256 GPUs. The run time statistics are collected using  \texttt{rocprof} during the training, and aggregated into 3 type of kernels, i.e., computation, communication (RCCL calls), and IO (data movements including device to host, host to device, and device to device). As expected, IO doesn't play a big role in LLM training on Frontier. ZeRO has the most data movements but the IO kernels take about 5\% of total run time. Communication becomes a bottleneck for training at scale, especially for larger models. For 6.7B with ZeRO stage 1, it accounts for about 40\%.} Note that in the above experiments, the per-device batch size is fixed. In the case of ZeRO stage 1, since the optimizer states are partitioned across all devices, the more GPUs, the less per-GPU memory footprint. Therefore, in practice, the per-device batch size can be increased to improve the scaling performance.

{\color{black}To better understand the profiling results, we plot the runtime and GPU power traces in Fig.~\ref{fig:profile}. These traces are collected using \texttt{OmniTrace}, and to avoid the excessive overhead, only one node runs the tracing. Since the workloads are evenly distributed, the resulted traces are representative for understanding the distributed training. One training step is shown in Fig.~\ref{fig:profile}, where the run time includes one forward and one backward step. The forward step consists of 32 layers (see MatGPT 6.7B architecture in Table~\ref{tab:arch}) operations, and each (zoom-in snapshot) is dominated by the flash attention operation (v2). In the backward step, the allreduce operation takes a significant amount of time, consistent with \texttt{rocprof} profiling (see Fig.~\ref{fig:scale}). The power traces (recorded per MI250X) shows high usage during the computation and drops down during the communication. All GPU traces behave similarly, confirming that the workloads are evenly distributed.    

{\color{black}In order to understand the performance of GPU computation, we investigated the breakdown of kernels that are executed within each transformer layer. The results are depicted in Fig.~\ref{fig:kernel-breakdown}, and demonstrate a few key takeaways. First, Fig.~\ref{fig:kernel-breakdown} (Left) shows that GEMMs account for the vast majority of a transformer layer's runtime, and their proportion increases with model scale ($65.9\%$ and $91.2\%$ for medium- and large-sized models, respectively). Second, Fig.~\ref{fig:kernel-breakdown} (Right) shows that within these GEMM kernels, the query-key-value (QKV) transformation in the attention layer along with the MLP layer account for the most runtime. Therefore, future optimizations targeting these blocks would benefit training time the most.}

\begin{figure}
\centering
\includegraphics[width=0.23\textwidth]{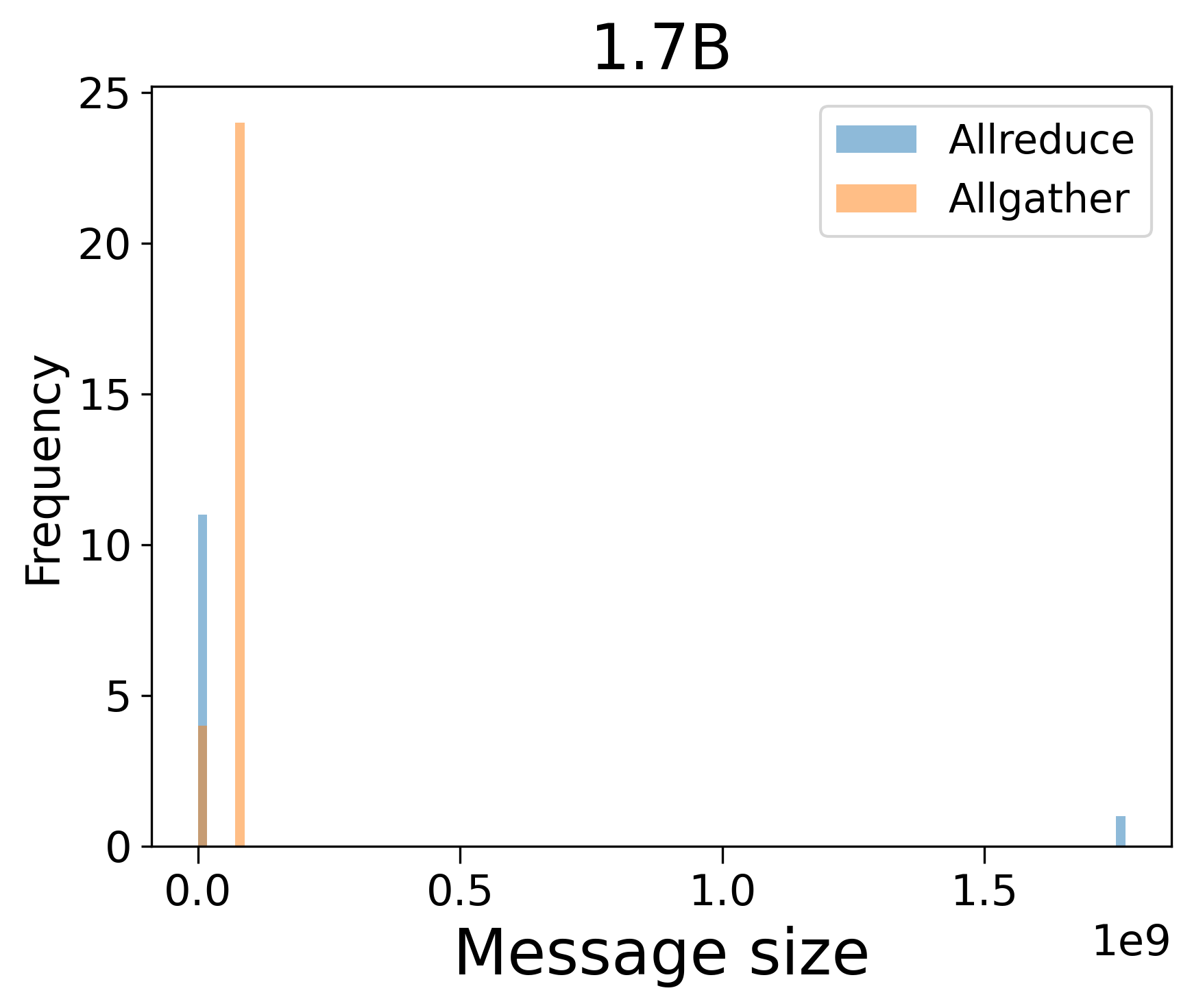}
\includegraphics[width=0.23\textwidth]{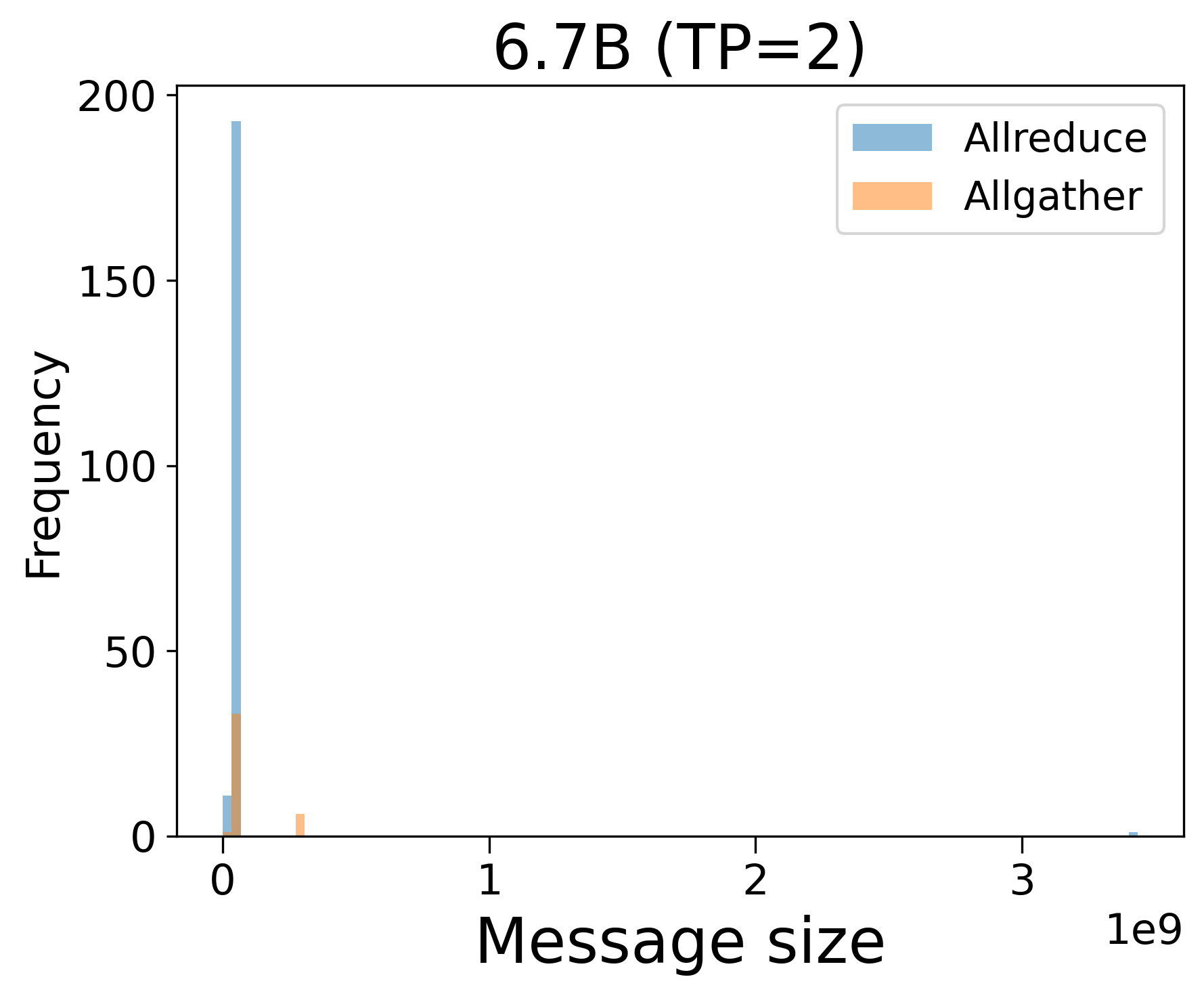}
\includegraphics[width=0.23\textwidth]{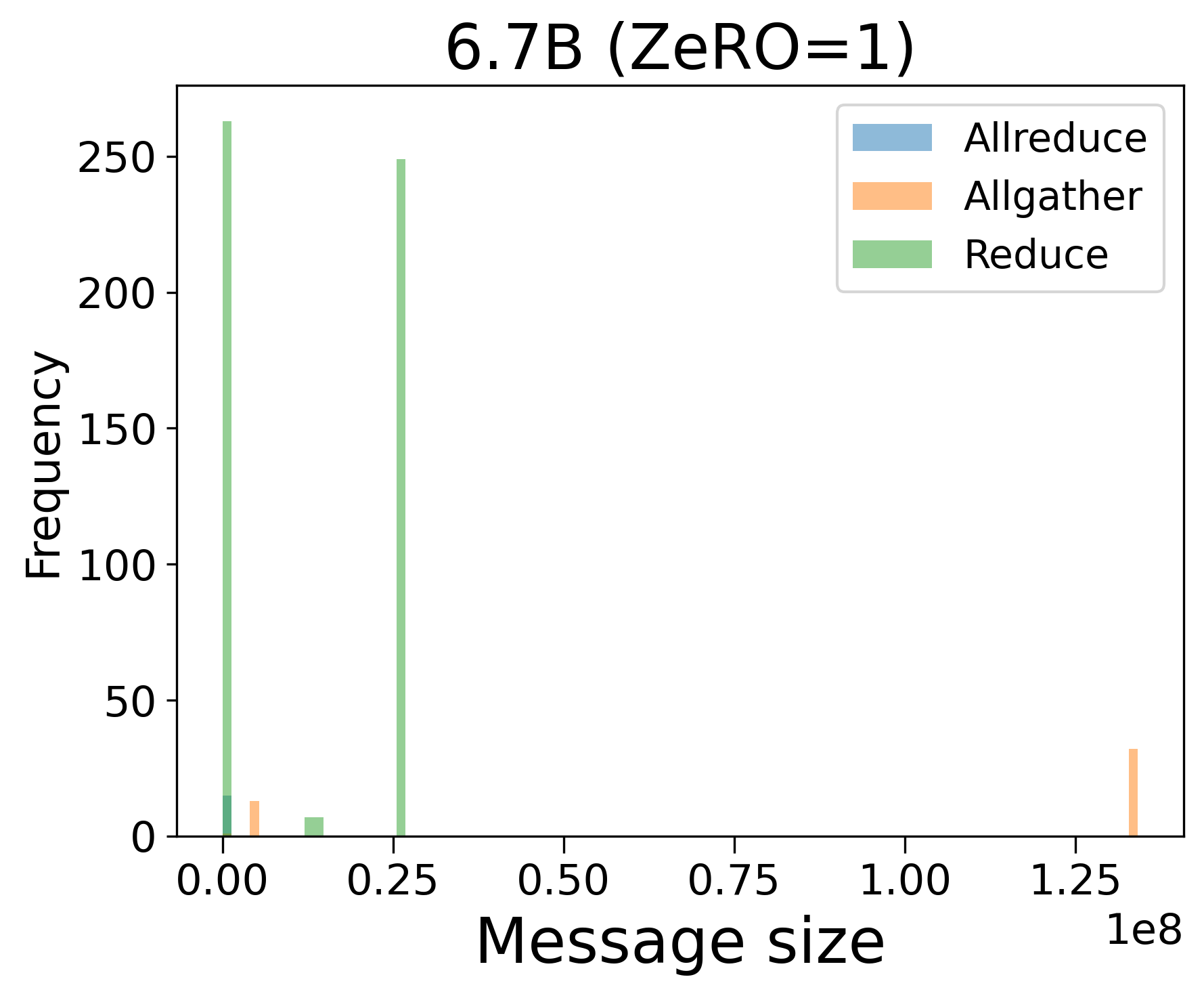}
\includegraphics[width=0.23\textwidth]{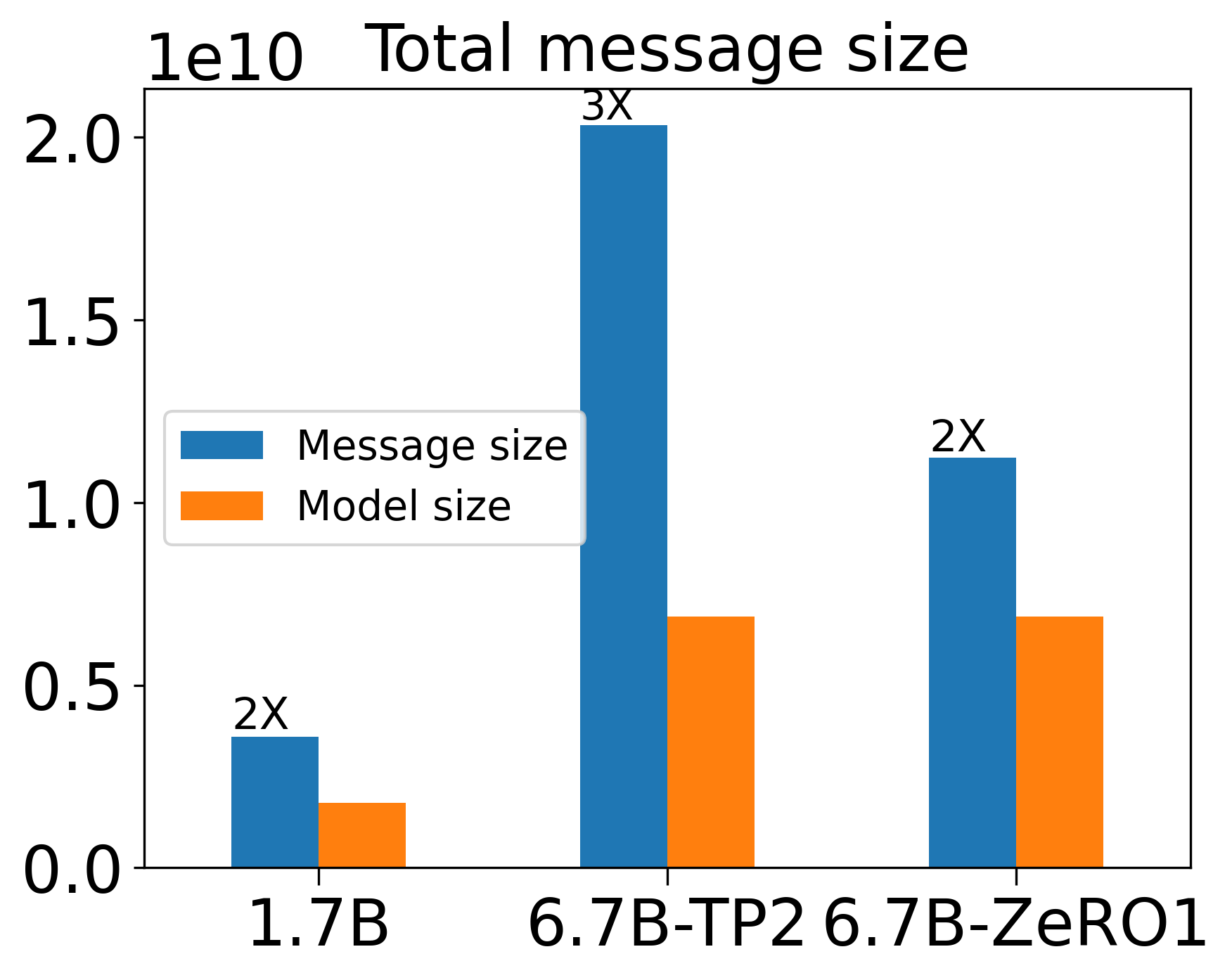}
\caption{\color{black}The histogram and aggregated message size per-batch-step per-GPU during the training of 1.7B with data parallelism, 6.7B with ZeRO stage 1, and tensor parallelism (TP=2), respectively.}\label{fig:msg}
\end{figure}

To further investigate the impact of communication, as shown in Fig.~\ref{fig:msg}, the histogram and aggregated message size per batch step per GPU are collected from RCCL logs (by setting \texttt{NCCL\_DEBUG\_SUBSYS=COLL}) for the three parallelism settings for the distributed training in Fig.~\ref{fig:scale}. The ZeRO stage 1 and tensor parallelism TP=2 for 6.7B incurs over an order of magnitude more RCCL calls (e.g., Allreduce and Allgather) compared to vanilla data parallelism for 1.7B. In terms of the total message size, both data parallelism and ZeRO parallelism (considered as a memory-efficient data parallelism) require a communication size about 2X the model size, while tensor parallelism requires 3X due to the additional communication of model parameters. Although TP=2 incurs a larger communication volume than ZeRO stage 1, the scaling efficiency is actually better because the 2 GCD within an MI250X has a 2X communication bandwidth compared to the inter-node communication needed for ZeRO stage 1 (see Fig.~\ref{fig:scale}).   }

\noindent {\bf Observation \circled{2}} For HPC platforms optimized {\color{black}for workloads with less demanding communication requirements}, adding extra parallelism dimensions such as tensor and pipeline usually adversely impacts the LLM training throughput. The recommended strategy is to keep model parallelism at the minimum and assign the rest of the computation resources to data parallelism. It is beneficial to map the partition of model parallelism to the platform network topology to maximize the the network bandwidth utilization.  

With the optimized computational performance, we pre-train MatGPT- NeoX and LLaMA on the full set of data tokens and compare the loss and the performance of downstream tasks in the following sections.

\noindent \textbf{\color{black}Power Usage and Cost Analysis}
\begin{figure}[t]
\centering
\includegraphics[width=0.5\textwidth]{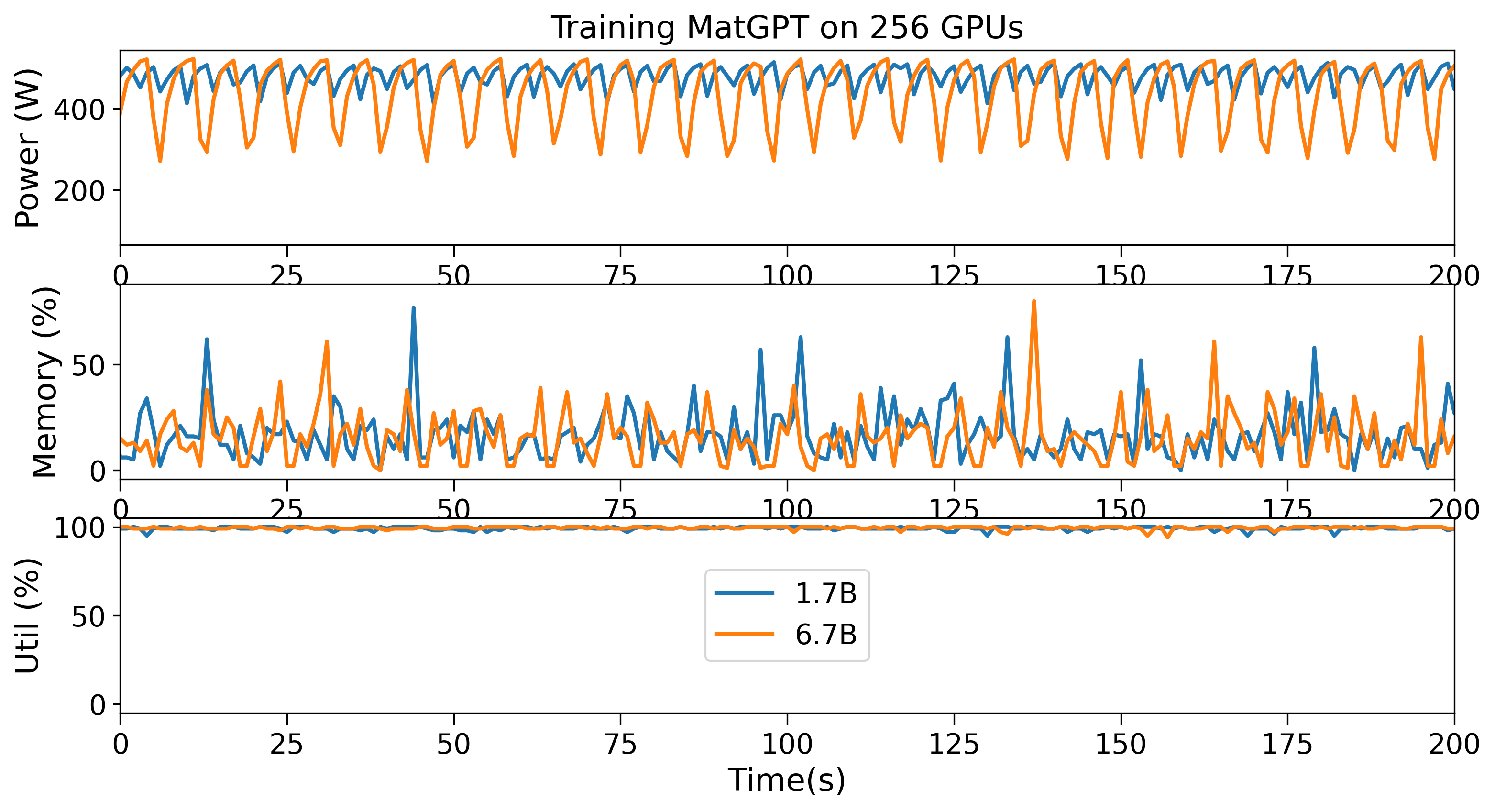}
\caption{The trace of power, memory, and GPU utilization for training MatGPT of 1.7B and 6.7B, respectively, with 256 GPUs on Frontier.}\label{fig:power}
\end{figure}
Pre-training LLMs is computationally expensive, and it's essential to be mindful of energy efficiency. We measure the power, memory, and GPU utilization during the training of MatGPT on Frontier {\color{black}with the \texttt{rocm-smi} tool. The update interval is per millisecond by default. Considering the duration of a training step is typically in seconds, \texttt{rocm-smi} can well capture the system metrics for most kernels during training. Due to the evenly distributed nature of the AI workload across all devices, the representative trace of one GPU is plotted}. As shown in Fig.~\ref{fig:power}, the power trace for training 6.7B shows larger oscillation than that of 1.7B, with a mean value of 434 and 476 W for 6.7B and 1.7B, respectively. Note that there is only one power sensor on an MI250X and the reported number is for the sum of 2 GCDs. Because the communication kernels also occupy GPU, the near 100\% GPU utilization for both cases is not a good indicator for the computation usage. Power actually correlates more closely with computational performance. {\color{black}In fact, the oscillation in power curves indicates the periodical computation and communication cycles, as shown in Fig.~\ref{fig:profile}, the power trace for a single training step.} Given the 75.9 and 80.5 TFLOPS per GPU for training 6.7B and 1.7B model with 256 GPUs, the energy efficiency for training can then be calculated as 0.27 and 0.33 TFLOPS/Watt, respectively. The training time and total energy consumption are also listed in Table~\ref{tab:power}. Note that the numbers are for training a single model, and in this study we have trained 6 models in total.       

\begin{table}[h]
\centering 
\caption{The time and energy usage for pre-training one 1.7B and 6.7B MatGPT model, respectively, on Frontier.} \label{tab:power}
\begin{tabular}{ccccc}
\toprule
Model & GPUs & \begin{tabular}[c]{@{}c@{}}Time\\ (hours)\end{tabular} & \begin{tabular}[c]{@{}c@{}}Energy\\ (MWh)\end{tabular} & \begin{tabular}[c]{@{}c@{}}Efficiency\\ (TFLOPS/Watt)\end{tabular} \\ \midrule
1.7B  & 256  & 4.1                                                    & 0.23                                                   & 0.33                                                               \\
6.7B  & 256  & 16.5                                                   & 0.91                                                   & 0.27                                                               \\ \bottomrule
\end{tabular}
\end{table}

\iffalse
\begin{figure}
\centering
\includegraphics[width=0.5\textwidth]{figures/scale.png}
\caption{The heatmap of training throughput (TFLOPS per GPU) for MatGPT with various number of layers and hidden sizes.}\label{fig:scale}
\end{figure}
\fi

\noindent \textbf{Loss Comparison}
It is shown \cite{scalinglaw} that the loss of LLM scales with the model and data sizes. The lower the loss, the better the model. Under the same experiment conditions, i.e., the same data processing and training pipeline, the loss can be used as an indicator to compare model performance. To identify the best model architecture on our 15B text corpus for materials science, we pre-train a suite of models on the Frontier supercomputer following a controlled recipe.

\begin{figure}
\centering
\includegraphics[width=0.24\textwidth]{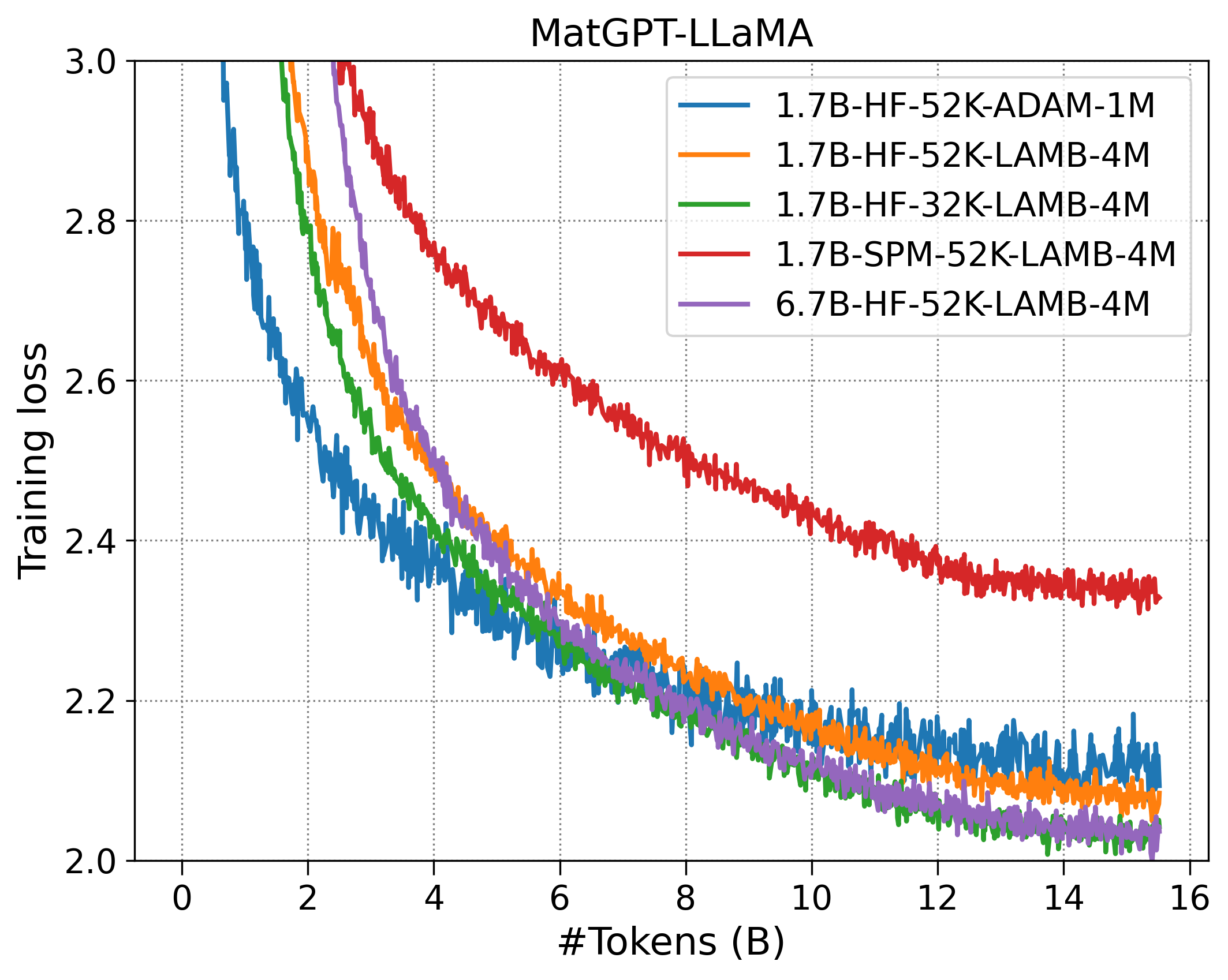}
\includegraphics[width=0.24\textwidth]{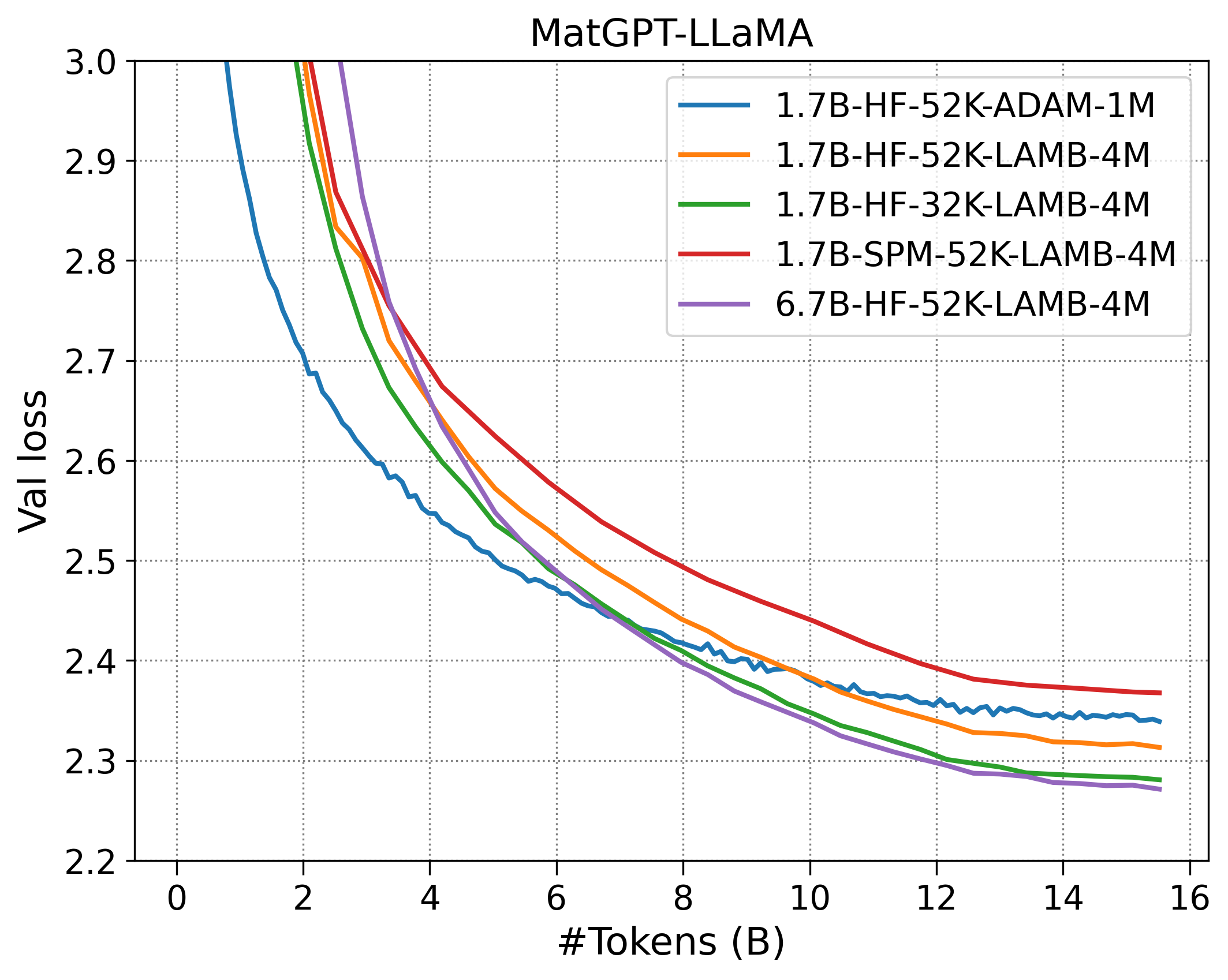}
\includegraphics[width=0.24\textwidth]{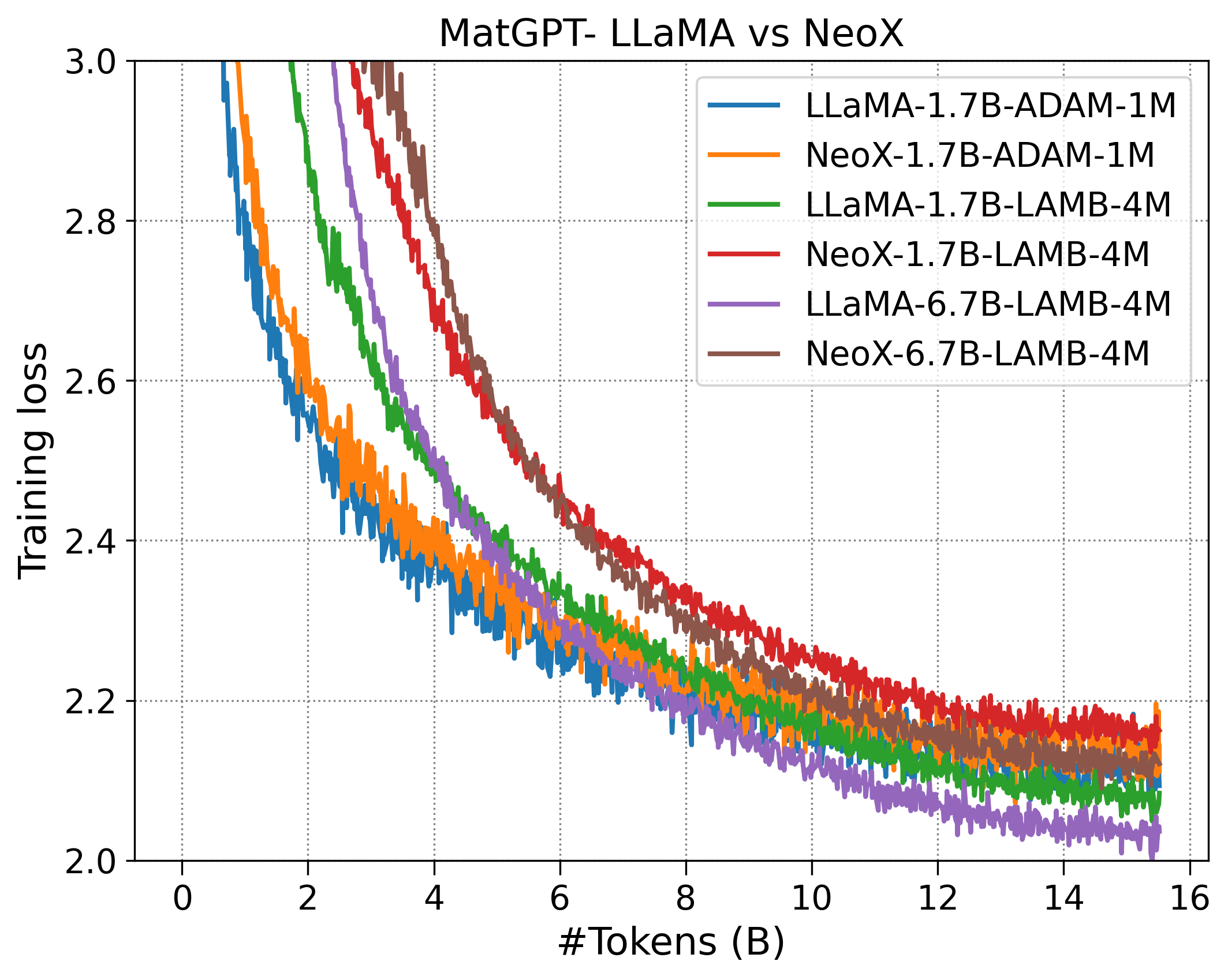}
\includegraphics[width=0.24\textwidth]{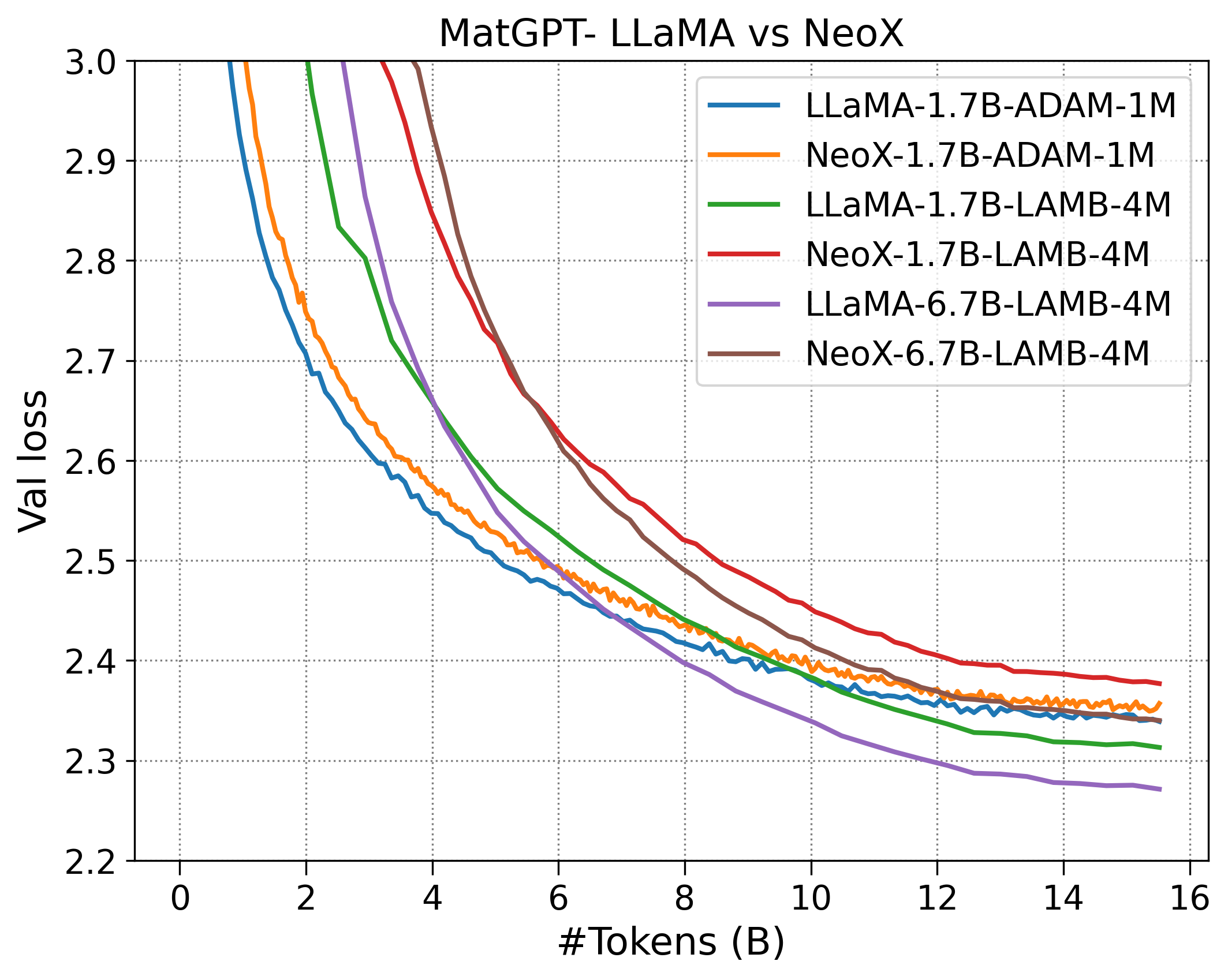}
\caption{\color{black}The training and validation losses of MatGPT-LLaMA models. Each curve corresponds to a pre-training experiment with the configuration of model size-tokenizer-vocabulary size-optimizer-batch size. Specifically, model size: 1.7B and 6.7B; tokenizer: HuggingFace (HF) and sentencepiece (SPM); optimizer: Adam and LAMB; batch size: 1M and 4M.}\label{fig:loss}
\end{figure}

In Fig.~\ref{fig:loss}, we plot the training and validation loss curves for all MatGPT-LLaMA architectures listed in Table~\ref{tab:arch}. As discussed in previous sections, to maximize the training throughput, most of the devices are assigned to data parallelism. This means that the training is performed with relatively large batch sizes. E.g., the GPT-3 model of a similar size\cite{gpt3} was trained with a batch size of 1M tokens using the Adam optimizer. Here we employ the LAMB optimizer to mitigate the generalization gap issue associated with large-batch training. By comparing the training and validation losses of the 1.7B model pre-trained on the same data (i.e., tokenized with the same HF tokenizer and 52K vocabulary size), the loss for the LAMB optimizer with 4M batch size is actually about 2\% smaller, indicating it's a better training procedure. Since the original LLaMA model used the SPM tokenizer with a vocabulary size of 32K, we train two additional 1.7B models using corresponding conditions while fixing the rest of the experimental configuration. In comparison, the loss is significantly bigger for SPM and much smaller for 32K. However, because the training data tokens are effectively different in these cases, the absolute value of loss cannot be compared. On the other hand, for the 6.7B model trained with the same data (i.e., HFT tokenizer, 52K vocabulary size), the loss is indeed smaller than that of the 1.7B model. {\color{black}We also explored the effect of the training precision, and found that the loss curves for MatGPT 1.7B, trained with float16 and bfloat16, are almost identical.} 

To compare the MatGPT- NeoX and LLaMA, in Fig.~\ref{fig:loss}, we plot the training and validation losses for corresponding 1.7B and 6.7B models. With the Adam optimizer and a batch size of 1M, the losses of both architectures are more or less the same. However, for the LAMB optimizer and a batch size of 4M, the MatGPT-LLaMA performs better for both model sizes.    

\noindent {\bf Observation \circled{3}} The losses for LLMs pre-trained with different tokenizers and/or vocabularies are not comparable. With the same pre-training recipe, the LLaMA architecture seems to provide a smaller loss than that of NeoX.

\iffalse
\begin{enumerate}
    \item performance heatmap 
    \item training loss: gpt-neox vs llama, hftokenizer 
    \item training loss: hftokenizer vs spm, llama
    \item accuracy progression on lm-eval 
    \item bar chart: downstream classification 
    \item bar chart: downstream regression 
\end{enumerate}
\fi

\noindent \textbf{Zero-shot Performance}
Although the LLM loss provides some indication of a model's performance, given LLM training is unsupervised, the downstream tasks are the ultimate metrics. Here we employ the popular question answering benchmarks \cite{eval-harness} including SCiQ \cite{Welbl2017CrowdsourcingMC}, PIQA \cite{bisk2020piqa}, OpenBookQA (OBQA) \cite{mihaylov2018can}, ARC-Easy (ARC-E) and challenge (ARC-C) \cite{Clark2018ThinkYH}, and Hendrycks colleague tests \cite{hendrycks2021measuring} on chemistry (HT-CC), physics (HT-CP), medicine (HT-CM), and compute science (HT-CCS). 

\begin{figure}
\centering
\includegraphics[width=0.24\textwidth]{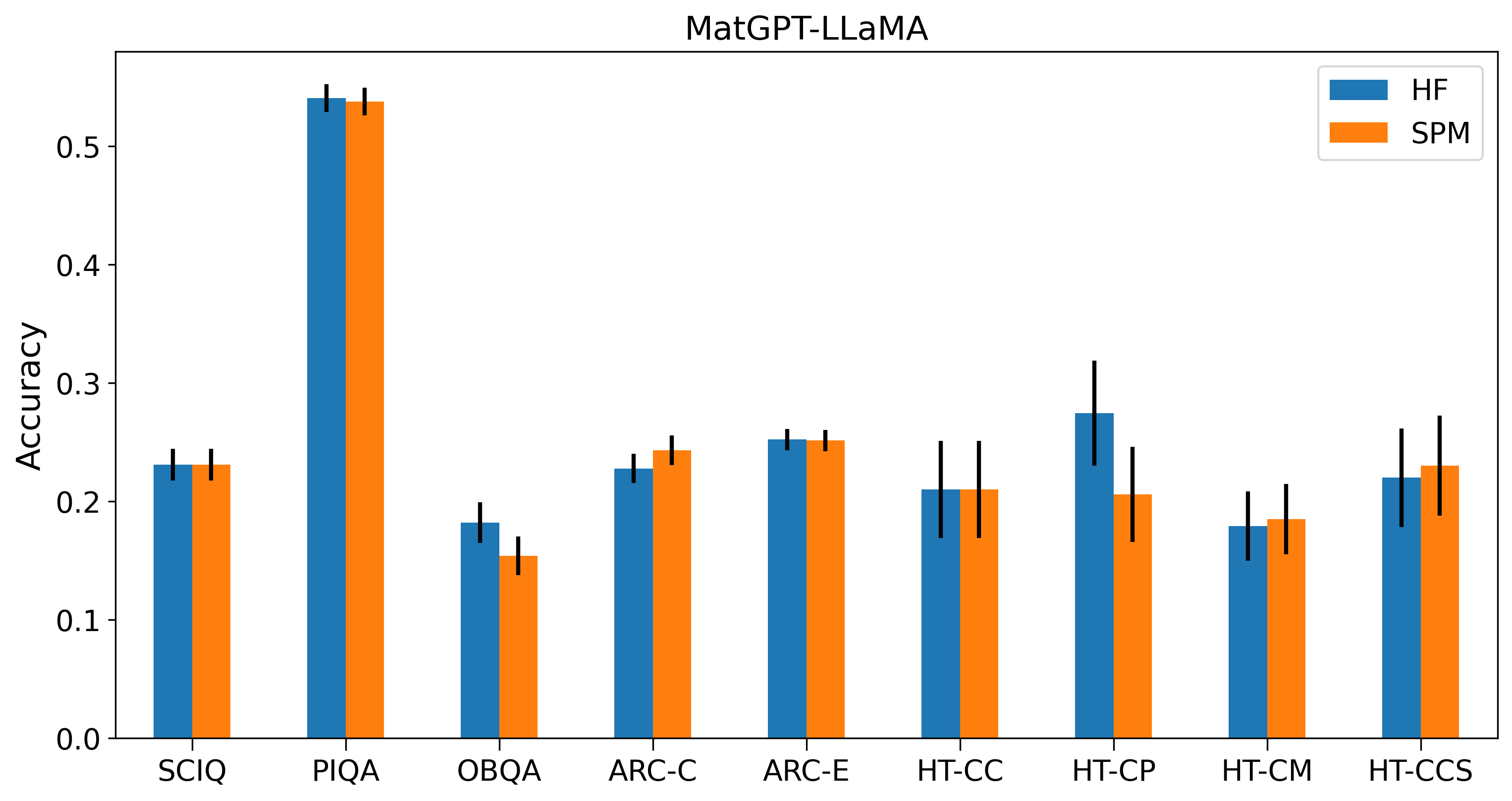}
\includegraphics[width=0.24\textwidth]{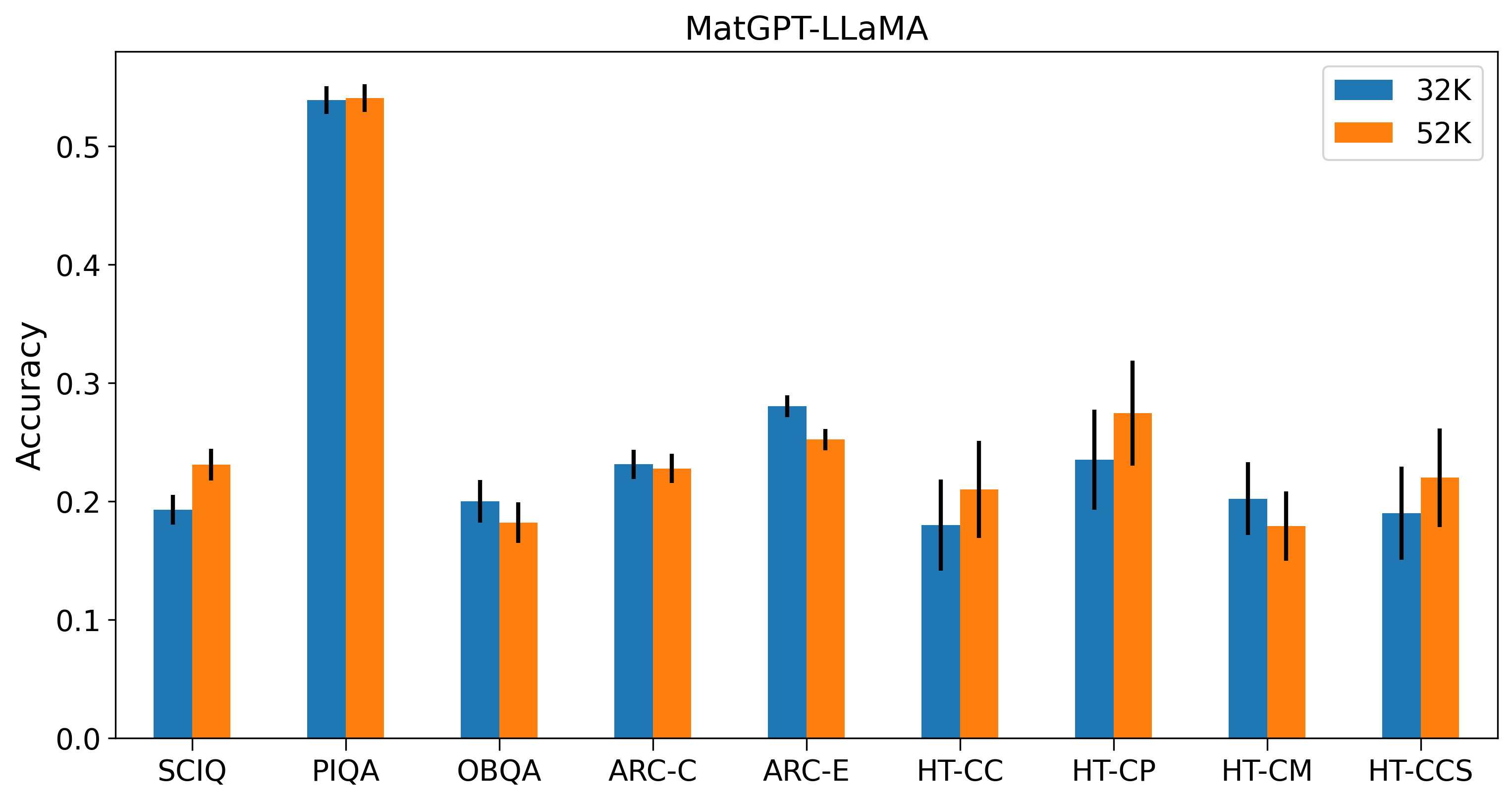}
\includegraphics[width=0.24\textwidth]{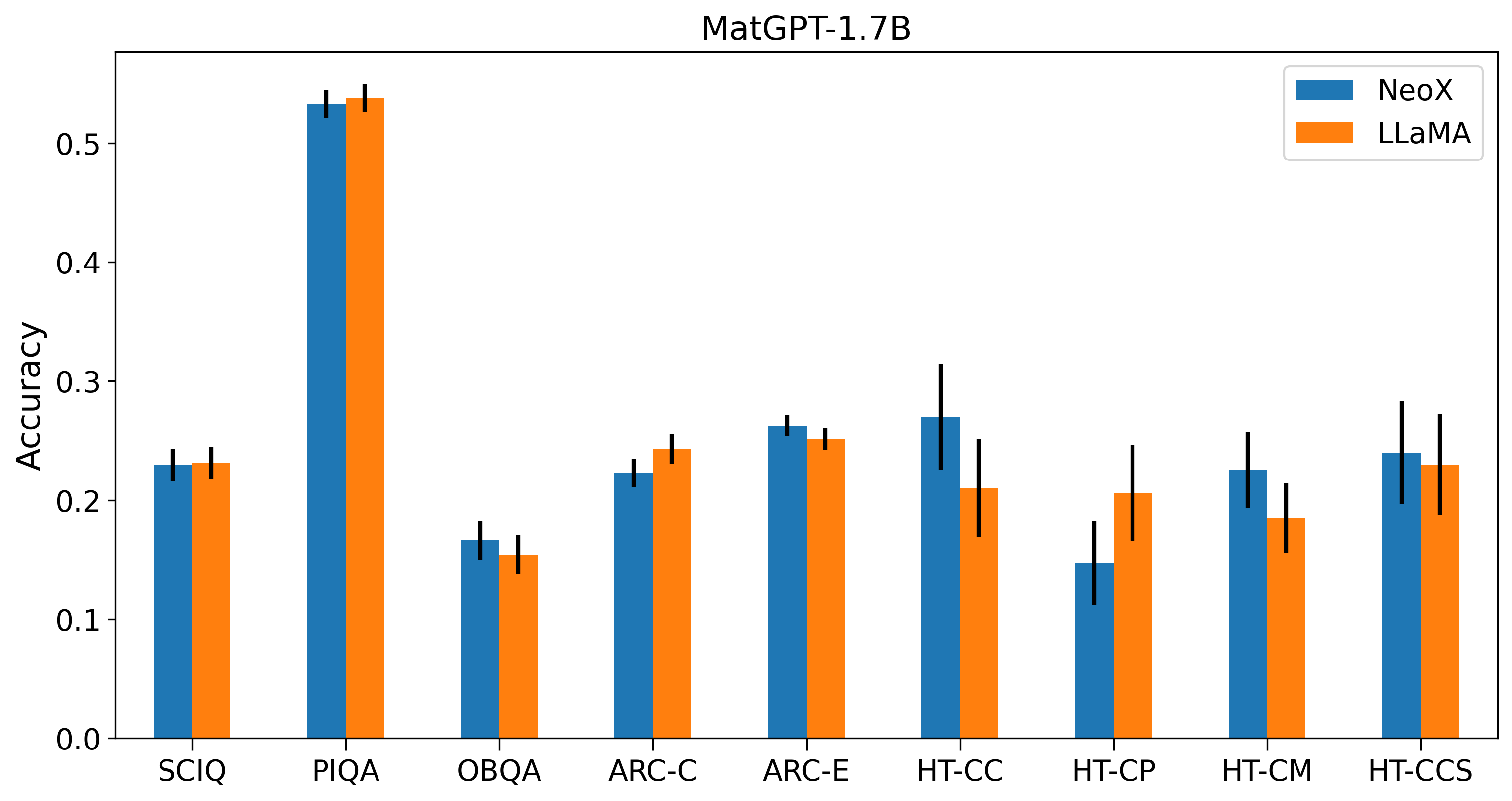}
\includegraphics[width=0.24\textwidth]{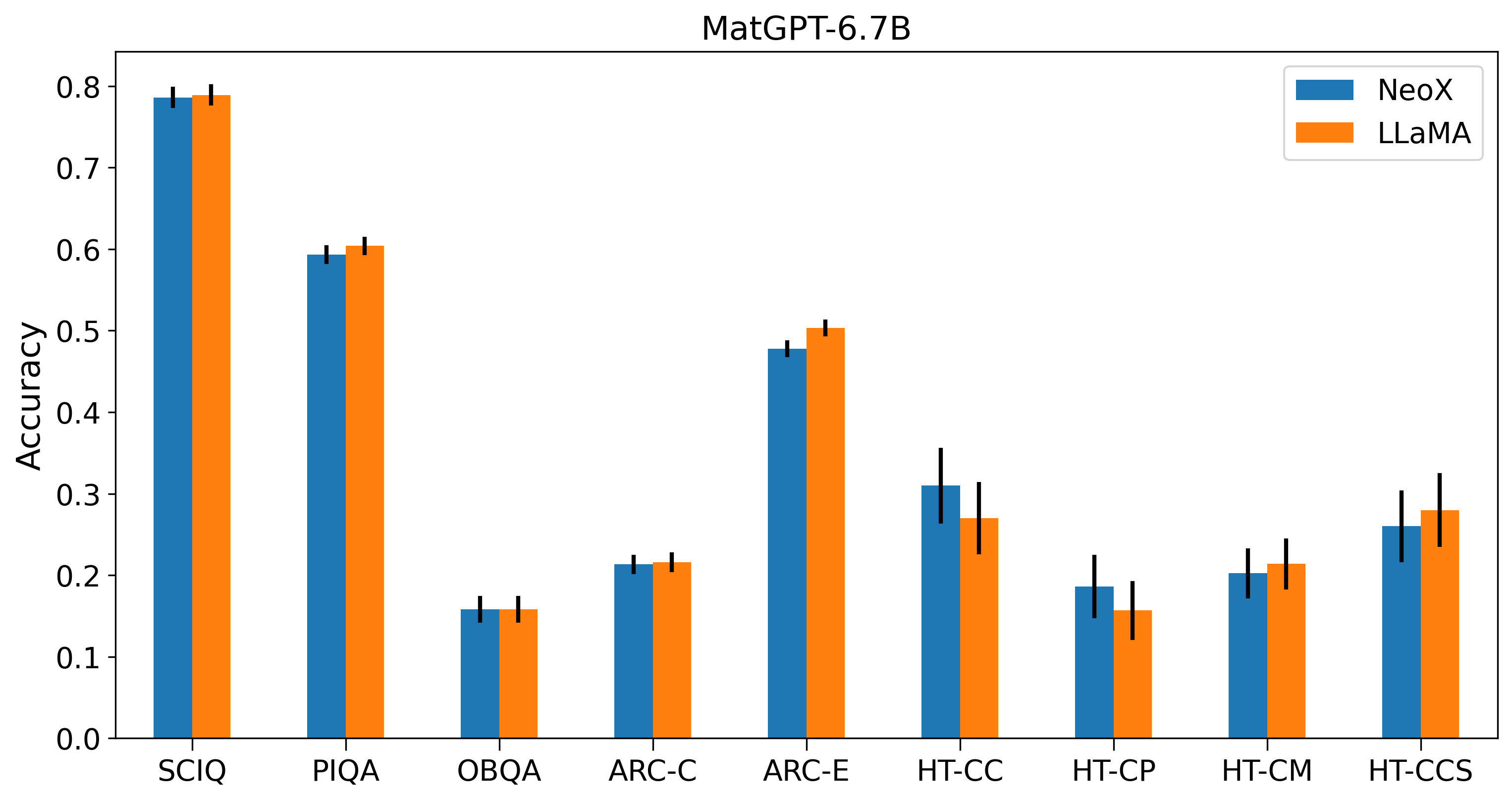}
\caption{\color{black}(Top) The zero-shot performance of MatGPT-LLaMA 1.7B models trained with HuggingFace (HF) and Sentencepiece (SPM) tokenizers on a vocabulary of size 52K, and HF tokenizer with 32K and 52K, respectively. The zero-shot tasks include SCiQ, PIQA, OpenBookQA(OBQA), ARC easy (ARC-E) and challenge (ARC-C), and Hendrycks colleague tests on Chemistry (HT-CC), Physics (HT-CP), Medicine (HT-CM), and compute science (HT-CCS). (Bottom) The zero-shot performance of MatGPT- LLaMA versus Neox models of sizes 1.7B and 6.7B.}\label{fig:llama-tokenizer}
\end{figure}

\iffalse
\begin{figure*}
\centering
\includegraphics[width=0.48\textwidth]{figures/1.7b-neox_vs_llama.png}
\includegraphics[width=0.48\textwidth]{figures/7b-neox_vs_llama.png}
\caption{The zero-shot performance of MatGPT- LLaMA versus Neox models of sizes 1.7B and 6.7B.}\label{fig:neox-vs-llama}
\end{figure*}
\fi

We start with the zero-shot test, where LLMs are directly applied to tackle new, unseen tasks. It can demonstrate a model's generalizability. In Fig.~\ref{fig:llama-tokenizer}, we show the effect of the tokenizer and vocabulary on the downstream language benchmarks. The zero-shot accuracies are plotted for the HF and SPM tokenizers with the same 52K vocabulary size, and the for 32K and 52K vocabulary sizes with the same HF tokenizer, respectively. The standard deviation in evaluating each benchmark is also plotted to show the variance in performance. The HF tokenizer seems to perform marginally better in 2 out of 9 tasks while the rest are about the same. For vocabulary size, 52K performs slightly better in 4 out of 9 tasks, while 32K shows an edge in 2 tasks. Considering our text corpus comes from scientific articles, larger vocabulary seems to be able to distinguish domain terminologies such as chemical elements in materials formulae.       

Using the same HF tokenizer and a vocabulary size of 52K, in Fig.~\ref{fig:llama-tokenizer}, we compare the zero-shot performance of MatGPT- NeoX and LLaMA models. For 1.7B models, NeoX performs marginally better in 4 out of 9 tasks while LLaMA is slightly better in 2 out of 9 tasks. For 6.7B models, LLaMA shows an edge for 2 tasks while the rest are on par. The results indicate that the loss (see Fig.~\ref{fig:loss}) does not fully correlate with the model's performance on downstream tasks.

\noindent \textbf{Few-shot Performance}
In addition to zero-shot tests, it is also important to evaluate LLMs' capability in adapting responses based on a few examples, i.e., few-shot performance. In Fig.~\ref{fig:fewshot}, we plot the 3-shot and 5-shot performance for MatGPT- NeoX and LLaMA of 6.7B parameters, respectively. For some tasks, e.g., SCiQ, prompting with more examples help to improve the model performance, and NeoX with 5-shot performs the best. Compared to zero-shot, the accuracy improvement for SCiQ is up to about 5\%. Overall, LLaMA performs slightly better in 3 out of 9 tasks while NeoX shows an edge in the other 3.     
\begin{figure}[]
\centering
\includegraphics[width=0.48\textwidth]{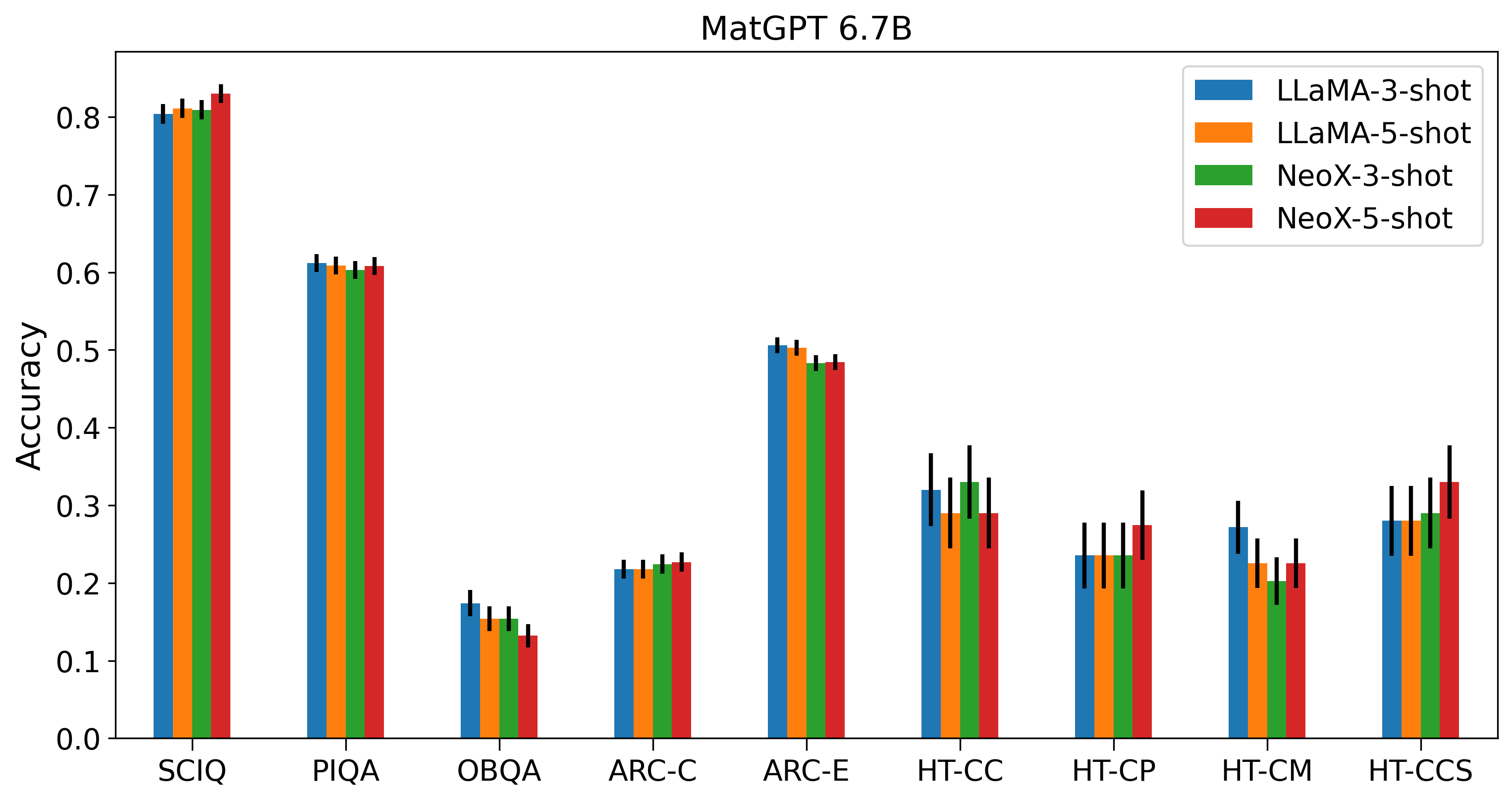}
\caption{The few (3 and 5) shot performance for MatGPT- NeoX and LLaMA. The zero-shot tasks include SCiQ, PIQA, OpenBookQA(OBQA), ARC easy (ARC-E) and challenge (ARC-C), and Hendrycks colleague tests on Chemistry (HT-CC), Physics (HT-CP), Medicine (HT-CM), and compute science (HT-CCS).}\label{fig:fewshot}
\end{figure}

\noindent {\bf Observation \circled{4}} The LLM loss can serve as an indicator of model performance but does not fully correlate with downstream task performance. GPT-NeoX and LLaMA architectures perform similarly on generic downstream tasks. Data quality seems to be the distinguishing factor, and a larger vocabulary size for scientific texts likely helps.  

\noindent \textbf{Fine-tuning for Scientific Task}
Beyond generic question answering tasks, more importantly, it is crucial to demonstrate the scientific benefit of a domain-specific LLM. Most of the efforts \cite{matscibert,matbert} so far have shown only the classification tasks, e.g., name entity recognition, but physical science is inherently numerical, with regression analysis serving as its core foundation. 
%For the regression task of band gap prediction, a BERT variant and a set of GPT variants are used in conjunction with a GNN architecture. 
Here we show that MatGPT can be used to improve the prediction quality of an important property of a material, i.e., band gap, which plays a key role in determining the material's electrical and optical properties. 

The experimental results support two complementary principles (i) LLMs of different architectures trained on larger datasets and a large number of parameters and (ii) models with higher hidden dimension sizes perform better than their simpler variants if these mentioned cases are true either individually or together. We first (i) present our band gap prediction results for different model settings where a set of domain-specific LLM variants (e.g. MatSciBERT, MatGPT-NeoX, etc.) is used in conjunction with a state-of-the-art GNN architecture, then (ii) show what intrinsic characteristics enable MatGPT embedding variants to perform better than each other and in general from MatSciBERT. Lastly, (iii) we will try to present evidence of the two characteristics mentioned above for various model architectures.

Table \ref{metric_eval} shows the performance of band gap prediction resulting from a GNN model used in conjunction with MatSciBERT and MatGPT material embeddings, respectively. In the table, CGCNN~\cite{xie2018crystal}, MEGNet~\cite{chen2019graph}, ALIGNN~\cite{choudhary2021atomistic}, and \mfnn~\cite{Cong_2023} are GNN implementations. ALIGNN and \mfnn represent state-of-the-art performance without pre-training, and +Scibert and +GPT represent \mfnn augmented with MatSciBERT and MatGPT embeddings, respectively. We observe 5\% and 8\% improvement with +SciBERT and +GPT, respectively, over \mfnn alone. As band gap prediction is extremely challenging, our result demonstrates the promise of leveraging LLMs in scientific applications.  

\begin{table}[htbp]
 \begin{center}
 \caption{Predicting band gap with various GNN implementations. Mean absolute error (MAE) is reported with the best value marked in bold.}
\label{metric_eval}
 \begin{tabular}{  c  c c  c  c c}
 \toprule
   CGCNN& MEGNet& ALIGNN& MF-CGNN& +SciBERT& +GPT\\  \midrule
   0.388&   0.33&     0.218&  0.215& 0.204& \bf{0.197}\\ \bottomrule
 \end{tabular}
\end{center}
\end{table}

 % With regard to the analyses, we separated the materials formula name into two sets, one for MatSciBERT and another for MatGPT, by comparing the predicted bandgap proximity to the ground truth while the embedding vectors are used with the GNN models.
\begin{figure}[h]
\centering
\includegraphics[width=0.24\textwidth]{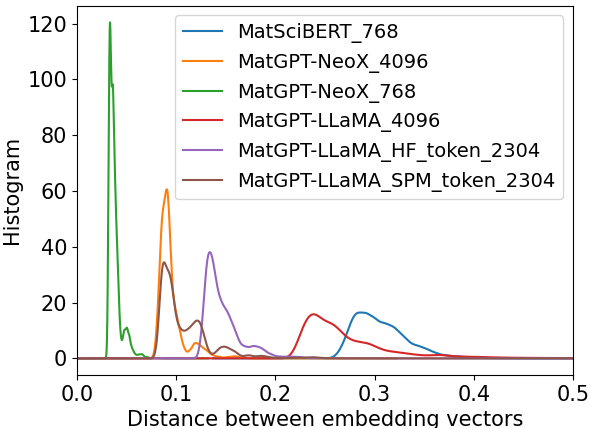}
\includegraphics[width=0.24\textwidth]{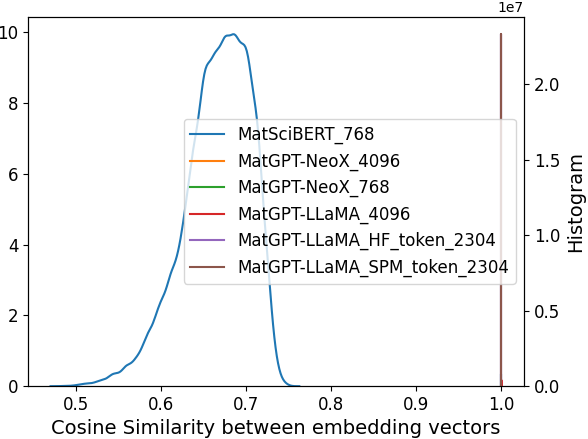}
\caption{\color{black}(Left) Euclidean distances between MatGPT embedding vectors of given material formulas. (Right) Cosine similarities between MatSciBERT embedding vectors of given material formulas.}
\label{fig:dist_comp}
\end{figure} 
To investigate the reason for obtaining results such as mentioned in table \ref{metric_eval}, we conducted fine-grained analysis on the MatsciBERT and MatGPT embedding vectors extracted via formula name of metal-organic materials (total 69240 materials) from respective pre-trained language model weights. The conducted analyses discover a couple of interesting facts from different viewpoints as follows that directly impact the regression results:
\begin{itemize}
    \item MatGPT embedding vectors are closer to each other than MatSciBERT vectors. This characteristic enables MatGPT variants to perform better in regression tasks than MatSciBERT. Fig.~\ref{fig:dist_comp} exhibits the density distribution of computed distances between embedding vectors for each LM pre-trained weight considered in this study where the histograms of all the GPT variants are located near the y-axis.
    \item All the MatGPT vectors point in the same direction meaning that the cosines between the respective embedding vectors of all GPT variants tend to be 1. This phenomenon is demonstrated in Fig.~\ref{fig:dist_comp} where density probabilities of cosine similarity between embedding vectors for all MatGPT variants seem to overlap on a vertical line. However, a different analysis demonstrates a model having smaller angles between MatSciBERT embedding vectors tends to lead to better performance than a model having larger angles between them. Although the embedding vectors of MatSciBERT and other GPT variants are derived from different architecture families, they both imply that smaller angles between embedding vectors perform better in bandgap prediction tasks.
    % This means vectors tend to point in the same direction and will eventually reduce their discriminative features to predict a regression value.
    % \item We apply the vector distance computation method for MatGPT performance analysis because MatGPT vectors point to the same direction so establishing a hypothesis on cosine similarity for regression performance comparison against the MatSciBERT would not be a reasonable method.
\end{itemize}

\begin{figure*}
\centering
  \begin{minipage}{0.3\textwidth}
  \subfloat[][MatSciBERT embedding clustering; \\Hidden size=768]{\includegraphics[width=1.0\linewidth]{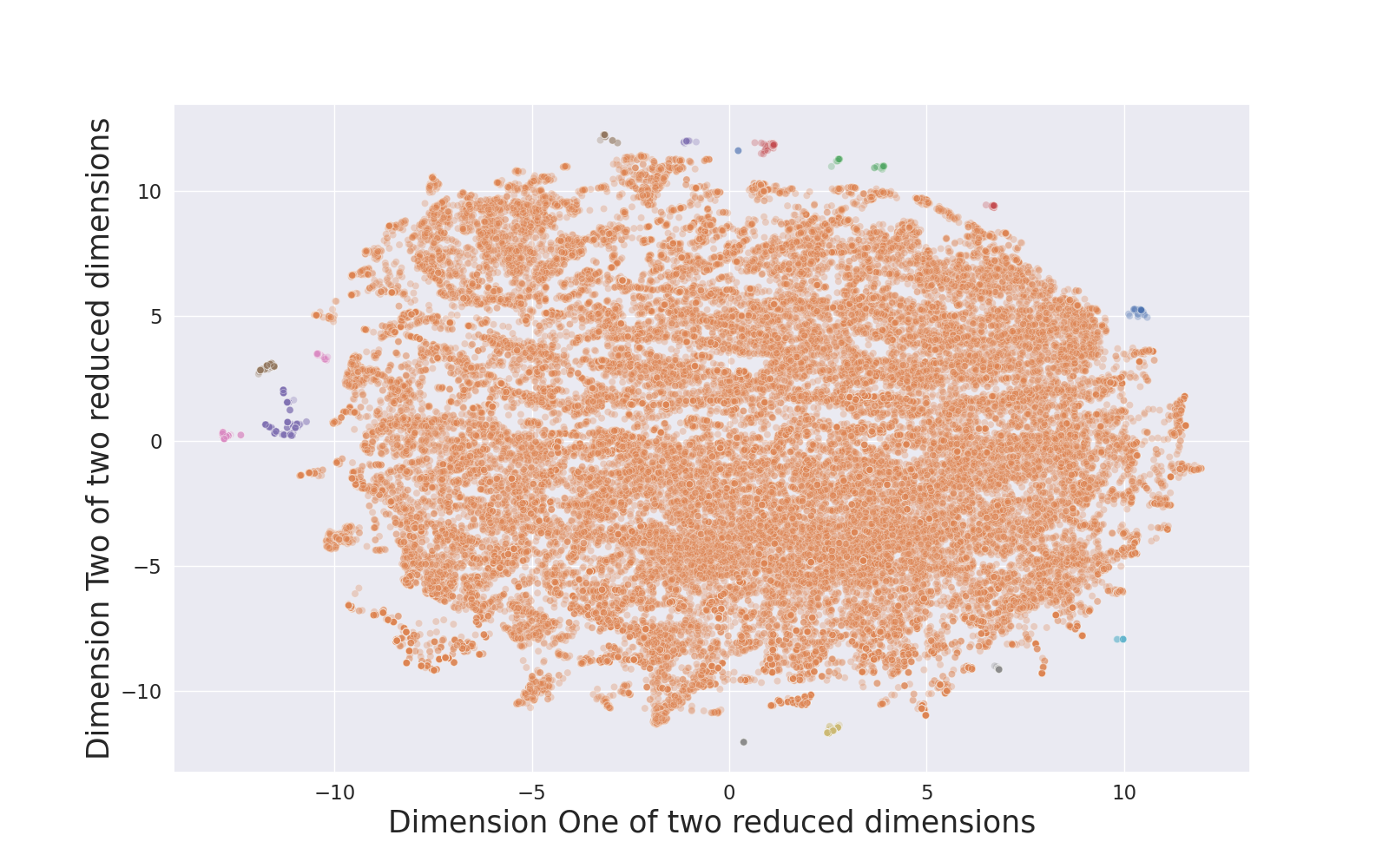}\label{bert768}}\\
  \subfloat[][MatGPT-NeoX embedding clustering; \\Hidden size=768]{\includegraphics[width=1.0\linewidth]{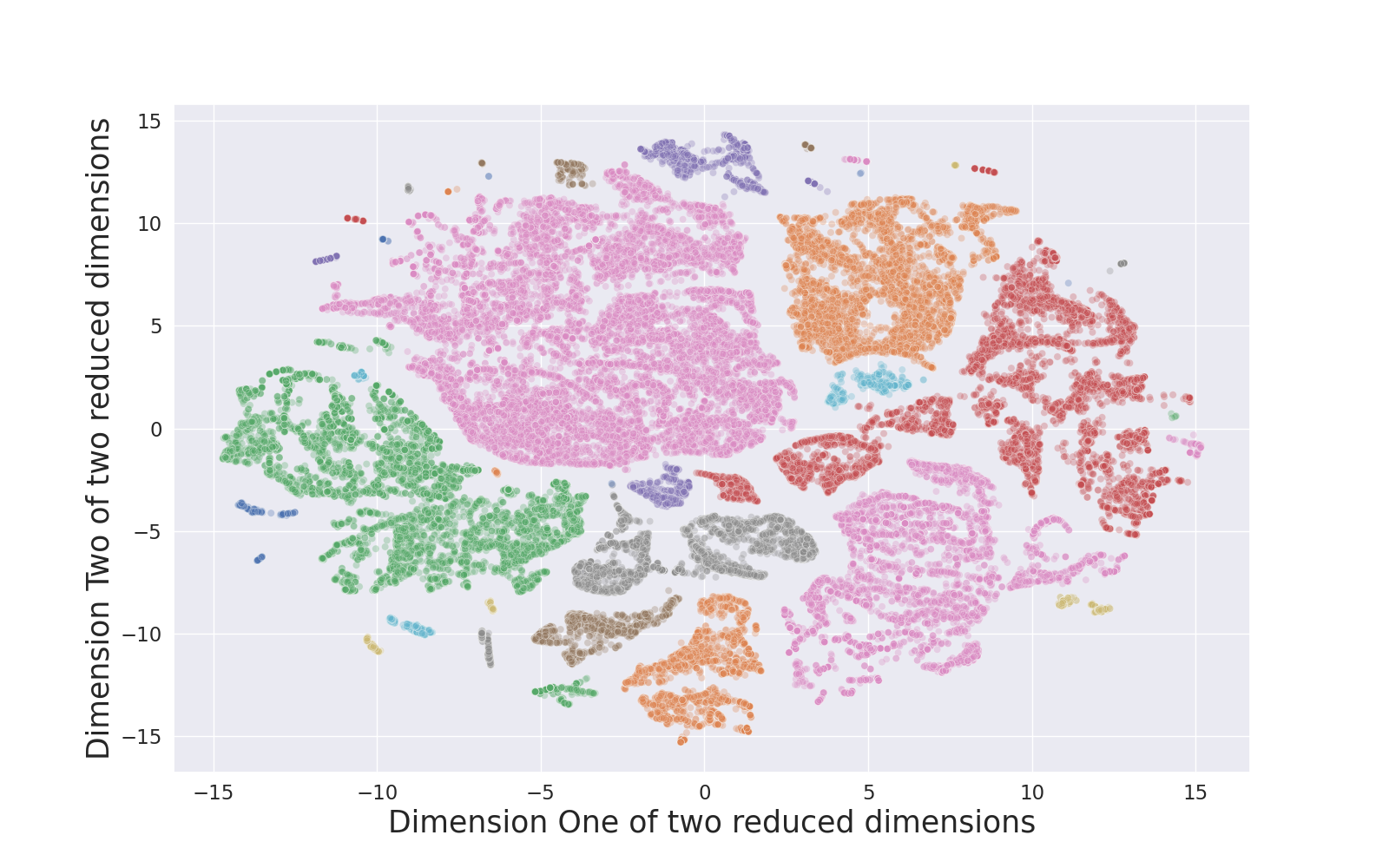}\label{gpt768}}  
  \end{minipage}
  %\subfloat[][GPT NeoX Embedding Clustering; Dimension size=2064]{\includegraphics[width=0.33\textwidth]{figures/gpt_2064_v2.png}\label{gpt2064}}
  \begin{minipage}{0.3\textwidth}
  \subfloat[][MatGPT-LLaMA 1.7B HF tokenizer \\embedding clustering; Hidden size=2304]{\includegraphics[width=1.0\linewidth]{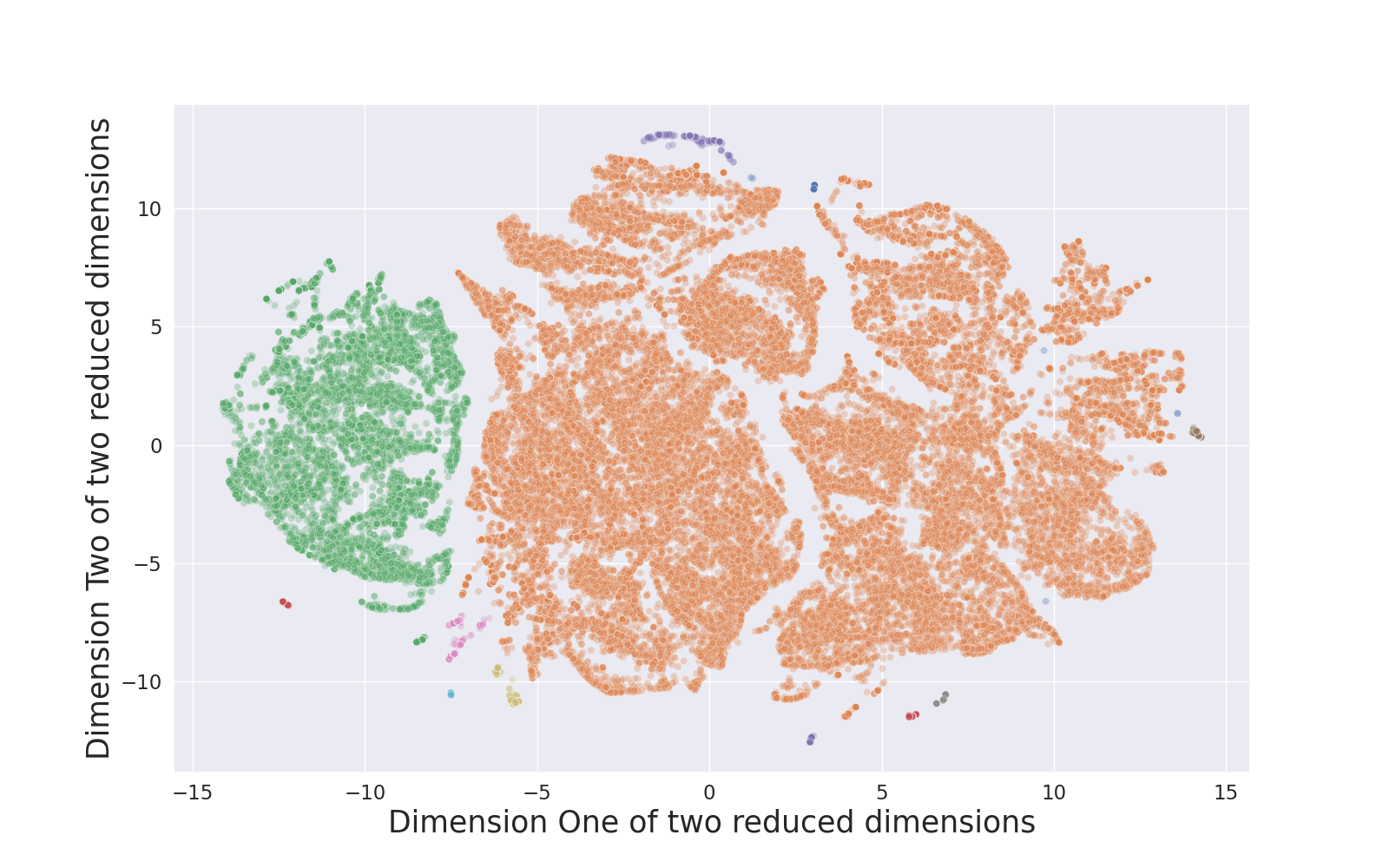}\label{LLaMA2304}}\\  
  \subfloat[][MatGPT-LLaMA 1.7B SPM tokenizer \\embedding Clustering; Hidden size=2304]{\includegraphics[width=1.0\linewidth]{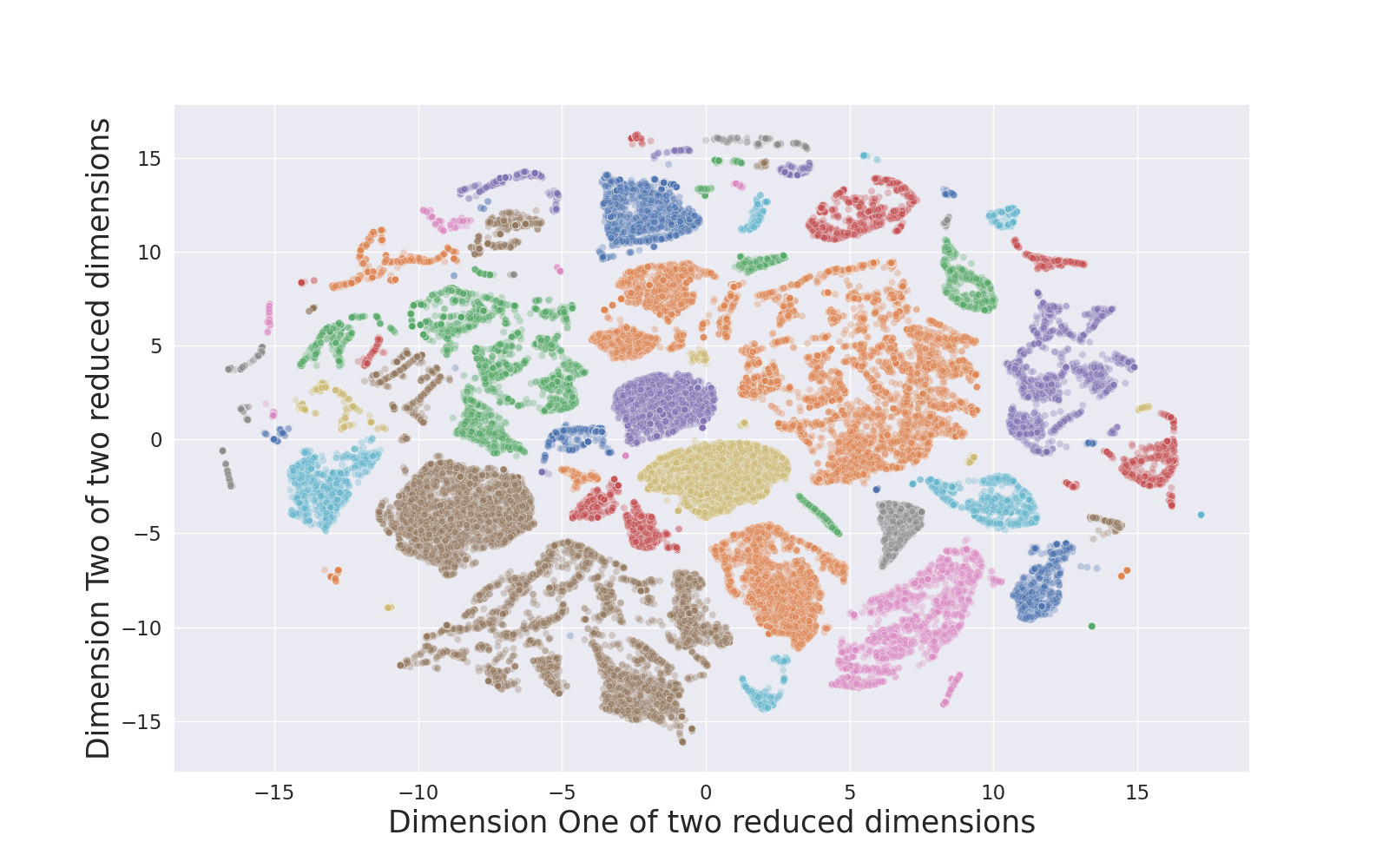}\label{LLaMAsp2304}}
  \end{minipage}
  \begin{minipage}{0.3\textwidth}
  \subfloat[][MatGPT-NeoX 6.7B embedding clustering; \\Hidden size=4096]{\includegraphics[width=1.0\linewidth]{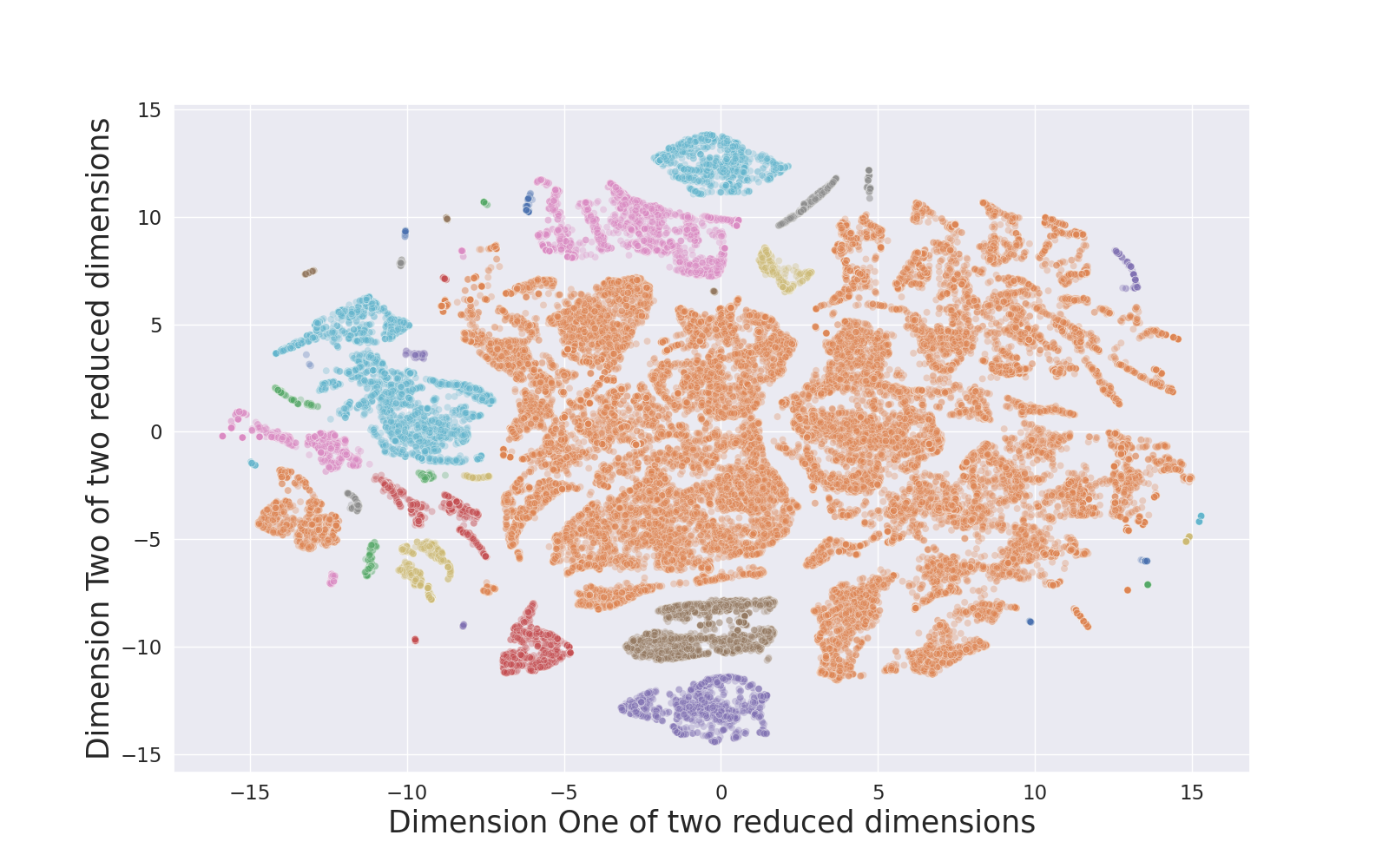}\label{gpt4096}}\\
  \subfloat[][MatGPT-LLaMA 6.7B embedding clustering; \\Hidden size=4096]{\includegraphics[width=1.0\linewidth]{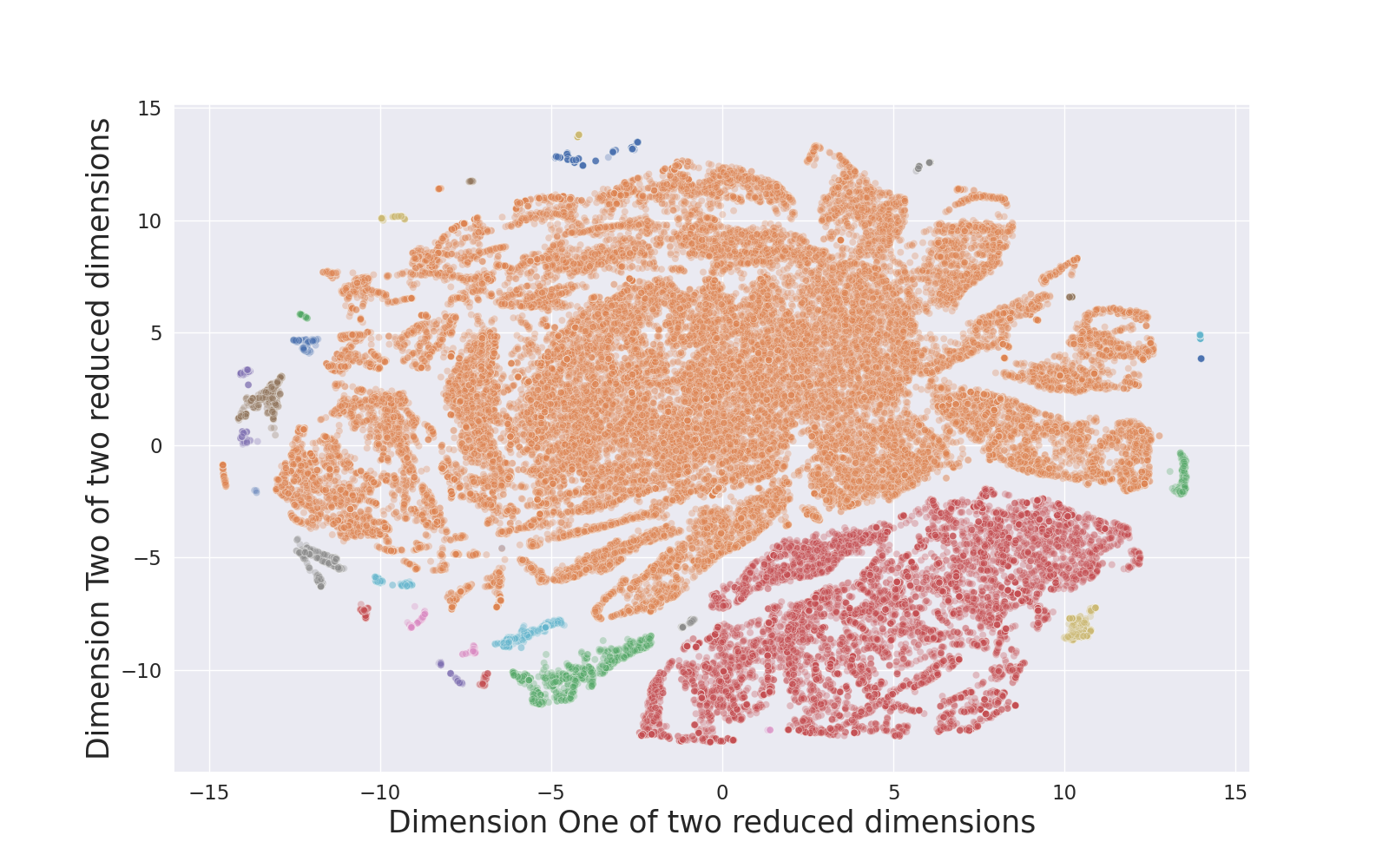}\label{LLaMA4096}} 
  \end{minipage}  
  \caption{Embedding clustering of material formulas for MatSciBERT and MatGPT variants after reducing dimensions by TSNE in tandem with PCA.}
  \label{embed_cluster}
\end{figure*}

To support our claim and establish the hypotheses mentioned above, we further investigate LM embedding spaces by demonstrating clustering plots generated from respective embedding vectors of formulas in Fig.~\ref{embed_cluster}. Compared to MatGPT-NeoX (Fig.~\ref{gpt768}) of similar size (hidden dimension of size 768), MatSciBERT embeddings form a very large cluster (Fig.~\ref{bert768}), which is an indicator of insufficient knowledge representation. Besides that, the embedding distances between MatSciBERT vectors are greater than the GPT variants' counterparts. This establishes a notion that data points are randomly disseminated in the low dimensional space, and hence perform a poor clustering. The optimal embedding clustering should reflect the characteristics of the prediction target --- band gap: materials in nature can be classified by band gap into a few categories, i.e., conductor, semiconductor, or insulator, and the band gap for each category of materials has its characteristics. It seems the knowledge embedded in MatGPT can serve as additional distinctive features to improve the GNN regression. Referring to the clustering principle that the same cluster data points are closer to each other than the data points in different clusters, embeddings of GPT variants maintain a pertinent balance between the overall embedding distances and comprised distinctive features. For MatGPT-LLaMA with different data tokenization (HF and SPM), the SPM (Fig.~\ref{LLaMAsp2304}) embeddings seem to overly classify formulas, and this is consistent with the performance comparisons for language benchmarks (See Fig.~\ref{fig:llama-tokenizer}).  

Our best-performing model (See Table~\ref{metric_eval}) is MatGPT-NeoX with a dimension size of 4096 plotted in Fig.~\ref{gpt4096} shows cluster results consistent with the prediction result. On the other hand, clusters of MatGPT-LLaMA embedding variants of different dimension sizes are plotted in Fig. \ref{LLaMA4096}, \ref{LLaMA2304}, and \ref{LLaMAsp2304}. They display either a higher number or a lower number of clusters, which seems to be the reason that LLaMA embeddings reduced performance as compared to NeoX.

\noindent {\bf Observation \circled{5}} For LLMs pretrained on scientific texts, model embeddings encode the knowledge of the literature. One risk-free usage of LLMs for science is the manipulation of embeddings for both classification and regression tasks.

\section{Conclusion} \label{sec:conclusion}
LLMs are poised to potentially revolutionize the way we conduct science and it is critical to establish best practices for deploying them on public HPC platforms, especially leadership supercomputers, to ensure the democratized usage of subsequent breakthroughs. In this study, we have systematically investigated two popular open-source LLM architectures --- GPT-NeoX and LLaMA. By designing controlled experiments, we carefully studied the effect of data tokenization, model architecure, and parameter count on both the behavior of training and validation losses, and the downstream language benchmarks. We outlined our observations and provided practical guidance for training LLMs on HPC systems. 

Furthermore, based on the scientific corpus we collected, we have pre-trained a suite of LLMs for materials science. We then demonstrated our suite on a scientific downstream application by injecting the models' embeddings into the graph neural network for fine-tuning, and achieved state-of-the-art performance for the band-gap prediction. 

Moreover, we reported the best-so-far LLM training performance on AMD GPUs, and demonstrated good scaling and energy efficiency on the Frontier supercomputer. The practical, end-to-end solution we establish can be applied to building LLMs on HPC systems in general.  

\section*{Acknowledgment}

This research was partially funded by a Lab Directed Research and Development project at Oak Ridge National Laboratory, a U.S. Department of Energy facility managed by UT-Battelle, LLC. This research used resources of the Oak Ridge Leadership Computing Facility (OLCF), which is a DOE Office of Science User Facility at the Oak Ridge National Laboratory supported by the U.S. Department of Energy under Contract No. DE-AC05-00OR22725.

\bibliographystyle{unsrtnat}
\bibliography{main}

\end{document}